\documentclass[journal]{IEEEtran}
\usepackage[shortcuts]{extdash}
\usepackage{bm,upgreek}
\usepackage{algorithm}
\usepackage{algpseudocode}
\usepackage{multirow}
\usepackage{bbm}
\usepackage{xcolor}

\usepackage{graphicx}
\ifCLASSINFOpdf
\else
\fi
\usepackage{amsmath}
\usepackage{amssymb}
\usepackage{amsfonts}

\usepackage{upgreek}

\MakeRobust{\Call}

\begin{document}
\title{Design and Implementation of Parametrized Look-Up Tables for Post-Correction of Oversampling Low-Resolution ADCs}

\author{Morriel~Kasher,~\IEEEmembership{Student~Member,~IEEE,}
        Michael~Tinston,~\IEEEmembership{Member,~IEEE,}
        and~Predrag~Spasojevic,~\IEEEmembership{Senior~Member,~IEEE}% <-this % stops a space

\thanks{This material is based upon work supported by the Office of Naval Research under Contract No. N68335-21-C-0625 and the National Science Foundation Graduate Research Fellowship under Grant No. DGE-2233066.}%
\thanks{M. Kasher and P. Spasojevic are with the Department of Electrical and Computer Engineering, Rutgers University, New Brunswick, NJ 08901 USA (email: morriel.kasher@rutgers.edu; spasojev@winlab.rutgers.edu)}%
\thanks{M. Tinston is with Expedition Technology, Inc., Herndon, VA 20171 USA (email: mike.tinston@exptechinc.com)}%
%\thanks{M. Shell was with the Department
%of Electrical and Computer Engineering, Georgia Institute of Technology, Atlanta,
}


%\markboth{IEEE Transactions on Signal Processing,~Vol.~??,~2025}%
\markboth{Preprint}%
{Kasher \MakeLowercase{\textit{et al.}}: Design and Implementation of Parametrized Look-Up Tables for Post-Correction of Oversampling Low-Resolution ADCs}



\maketitle

% \begin{abstract} % required: 150-250 words
% We propose a novel framework for the design, optimization, and implementation of Look-Up Tables (LUTs) used to recover noisy, oversampled, quantized signals given a parametric input model and prior distribution for its parameters. The proposed methodology decomposes the intractable LUT design problem into four distinct stages, each of which can be addressed analytically using a model-driven approach without reliance on training from any experimental or simulated dataset. Two novel indexing schemes are proposed to limit the LUT memory overhead while three dithering methods are studied to improve a spectral purity metrics. The resultant LUT is used to emulate the spectral effects of pre-quantization dithering through an all-digital solution applied after quantization. This new LUT design is tested with a noisy sinusoidal input quantized to 3 bits and shown to improve its Spurious-Free Dynamic Range (SFDR) by over 19 dBc with only 324 bytes of memory while maintaining the same 3-bit fixed-point precision at the digital output. This correction can be implemented in a two-level combinational logic for ultra-low latency and, hence, feasibly supporting wideband low-resolution devices.
% \end{abstract}
\begin{abstract} % required: 150-250 words
We propose a  framework for the design, optimization, and implementation of Look-Up Tables (LUTs) used to recover noisy, oversampled, quantized signals given a parametric input model.
%and prior distribution for its parameters. 
The LUTs emulate the spectral effects of pre-quantization dithering through an all-digital solution applied after quantization. This methodology decomposes the intractable LUT design problem into four distinct stages, each of which is addressed analytically using a model-driven approach without reliance on training.
%from any experimental or simulated dataset. 
Three dithering methods are studied to improve spectral purity metrics. Two novel indexing schemes are proposed to limit the LUT memory overhead shown to compress the LUT size by over four orders of magnitude with marginal performance loss. The LUT design is tested with an oversampled noisy sinusoidal input quantized to 3 bits and shown to improve its Spurious-Free Dynamic Range (SFDR) by over 19 dBc with only 324 bytes of memory while maintaining the same 3-bit fixed-point precision at the digital output. This correction can be implemented using two-level combinational logic ensuring ultra-low latency and, hence, suitable for low-resolution wideband devices.
\end{abstract}

\begin{IEEEkeywords}
Analog-to-digital conversion, quantization, look-up tables, dithering.
\end{IEEEkeywords}

% \fi
\IEEEpeerreviewmaketitle

\section{Introduction}
\IEEEPARstart{Q}{uantization} 
% why quantization
(or analog-to-digital conversion) is a ubiquitous process in audio/video, measurement, data compression, and communication systems. A quantizer/analog-to-digital converter (ADC) applies a hard non-linearity to a continuous-domain analog input to produce a discrete-valued output. While this operation is necessary to represent signals digitally or reduce their size, it introduces quantization error which distorts the input signal and limits the accuracy of its digital representation \cite{GrayQuant1998}.

% dithering fails bc hard to implement
Conventional quantization produces an error process that is highly (self-)correlated and also correlated with the input signal, resulting in prominent quantization artifacts that can reduce the perceived fidelity of quantized data~\cite{GrayDither1993}. This effect is especially significant in low-resolution quantization required for low-latency wideband applications and, hence, motivates a method to decorrelate quantization error.
%from the input samples. 
%for coarse quantizers. 
The non-subtractive dithering method achieves this by adding an analog dither signal to the input prior to quantization~\cite{wannamaker1997thesis}. When drawn from an appropriately-chosen distribution, the dither signal can render conditional moments of the error process independent of the input~\cite{WannamakerNSD2000,vanderkooy1984resolution}. This approach has been shown to significantly improve perceptual quality of the quantized output~\cite{WannamakerPsychoacoustics1992}.

% LUTs, good but non-analytical besides MSE
While dithering is an attractive and effective option for decorrelating the quantization error, it can be challenging to implement in practice. This is because it requires real-time generation and addition of the dither analog signal 
%from a pre-defined distribution 
prior to analog-to-digital conversion, typically necessitating an entire parallel digital-to-analog converter (DAC) chain. 
%Moreover, ADCs or digitizers often exist as fixed units where the analog input side is inaccessible for modification after a chip has already been manufactured. 
%Hence, addition of an analog dither signal to the input of an existing ADC may be either 
%Hence, this approach may be impractical or outright infeasible. 
In this paper we consider a method for compensating ADC quantization error exclusively in the digital-domain (i.e., post-quantization). 

A Look-Up Table (LUT) is an all-digital post-processing method used to improve quantizer performance efficiently. A state-space-indexing LUT uses $N$ previous quantized output values to index a correction value which then replaces the current digital output~\cite{lundinthesis,lundin_characterization_2005}. $N$  is the order or dimensionality of the LUT. It can be implemented using  two-level combinational logic circuits having $\mathcal{O}(1)$ processing time complexity, 
%whose speed is only limited only by the propagation delay of two series logic gates. This 
making it well-suited for wideband applications due to its extremely low latency. Furthermore, as an all-digital method, it can be 
%designed to output in the same fixed-point precision as its input, making it plug-and-play compatible 
integrated with an existing digital signal processing back-end.

% Despite promising characteristics for post-ADC correction,
Nevertheless, existing work on LUT design for pos-ADC correcion is limited. The choice of LUT output value is non-trivial. Prior work optimizes its design for Mean Square Error (MSE)~\cite{adctextbook}, Total Harmonic Distortion (THD)~\cite{lundin2003minimalthlut,hummels1,hummels2}, and Spurious-Free Dynamic Range (SFDR)~\cite{kashersfdrlut}. 
However, these works lack a set of cohesive LUT design principles--- existing literature relies almost exclusively on data-driven calibration procedures to train and optimize the LUT entries numerically~\cite{hummels1,de_vito_bayesian_2007,lundin_external_2001,gines_digital_2021}.
These design methods can be unreliable (subject to the quality of training data), uninformative (functionally black-box approaches), and may have unpredictable performance (difficult to be studied analytically).
%However, work on the latter two metrics relies on training (calibration) of LUT entries rather than direct analytical computation. 
The only analytically-derived LUTs are 
%The advantage of LUTs is their low implementation complexity, high efficiency (indexing requires no computation and thus allows real-time correction), and entirely digital application. Their primary disadvantage is size, as the memory required to store all of the values in a LUT is exponential with both its order $N$ and resolution $b$ such that the memory size in bits is $b \cdot 2^{bN}$. Performance improvement of low-order LUTs is highly limited, and high resolution often necessitates lower LUT order to minimize memory size.
%Since a high order is often necessary to achieve good performance, LUTs are typically only deemed practical for low-resolution quantizers.
% LUTs
the popular Midpoint and MMSE LUTs, designed only for $N=1$~\cite{adctextbook,attivissimomidpointlinearizationNSdither,lundin_bounds_2009}. These works restrict LUTs to compensating ADC irregularities (e.g., for a non-uniform transfer function), and they typically require high-precision outputs, whereas, they suffer severe performance penalties when their resolution is limited to that of the input~\cite{lundin_adc_2005}.
%severely limits its performance capabilities but is a highly desirable property for the system to maintain. % cite lundin supra-bit precision here
Another important research direction are parametrized LUTs. One popular example are the frequency-selective LUTs~\cite{lundin_analog--digital_2002}, where the input signal is assumed to be a  tone and its  frequency is estimated using  an additional LUT~\cite{andersson_frequency_2000,kasherfreqestlut}.
Moreover, all existing LUT-based corrections require large memory size that grows exponentially with $N$. Such LUTs arecimpractical for memory-constrained systems such as Field Programmable Gate Arrays (FPGAs) and Integrated Circuits (ICs).

% now we combine all three analytically, where our proposed method is actually a supserset of all prior LUTs
To overcome these limitations, we study a broad class of all-digital model-based  LUTs which we term \textit{dithered parametrized look-up tables}. 
Proposed design consists of an indexing, an estimation, a dithering, and a re-quantization stage.
Prior information about the input signal informs its parameters estimated by a LUT indexed using quantized low resolution ADC samples. {\em Masked indexing} techniques  support drastic  reduction of the LUT memory size. %The objective of these LUTs is to accurately recover input signals after low-resolution noisy quantization, a more generalized application than traditional ADC post-correction. 
% One popular related example is the frequency-selective LUT described in \cite{lundin_analog--digital_2002,} where the tone parameter is its frequency which is estimated using  an additional LUT \cite{andersson_frequency_2000} \cite{kasherfreqestlut.} 
% %In our prior work, we 
Several strategies emulate the desirable effects of dithering in the digital domain, an approach we term {\em post-quantization dithering}. % This paper is an in-depth treatment of that initial proposal. 
They allow the designer to trade MSE with SFDR via digital randomization, which reduces error correlation and improves spectral purity  at the cost of increased noise power~\cite{kasher_postquantization_2024}. Re-quantization to low/original resolution ensures that the low-latency and wideband post-processing are still feasible 
%post-correction 
while maintaining performance. 
The proposed post-correction system possesses several distinct features and advantages, some of which we demonstrate in this work and summarize next:
\begin{itemize}
    \item All the benefits of LUT-based post-correction methods including: all-digital implementation, $\mathcal{O}(1)$ access time complexity, two-level combinational logic for ultra-low latency, and pre-computation of entry values allowing training algorithms of arbitrary computational intensity without impact on run-time performance.
    \item Signal recovery following low-resolution quantization, in spite of strong deterministic non-linear distortion that cannot be well-approximated by 
    %any conventional stochastic noise models such as 
    pseudo-quantization~\cite{widrow_statistical_1996} or additive noise~\cite{marco_validity_2005}.
    \item A design having the same fixed-point output precision as input samples, supporting plug-and-play integration with any existing digital back-end by maintaining the same digital throughput.
    %\item Emulation of analog dithering through a parametric digital dithering block, allowing future application-specific or real-time LUT optimization by controlling the dither distribution without changing any of the pre-computed LUT entries.
    \item Novel methods of digital dithering within a LUT correction structure.
    \item LUT indexing schemes that can reduce memory requirements by several orders of magnitude.
    \item Tractability of the LUT design via its decomposition into several  model-driven design stages, each analytically optimized, without reliance on traditional data-driven calibration  or numerical optimizations.
    \item For a 3-bit quantized sinusoidal input we demonstrate an improvement in MSE by $>9$ dB with $1446$ bytes of memory and an improvement in SFDR by $>19$ dBc with only $324$ bytes of memory.
\end{itemize}
%The system can also support several auxiliary benefits not studied in this work directly, including: versatility to arbitrarily non-linear quantizers, effective in-band denoising for wideband inputs that outperforms optimal linear filtering techniques, and robustness to simultaneous interference signals at the quantizer input.
The proposed low latency post-correction  approach makes  an attractive option for wideband digitizers, where input signals are typically highly oversampled. Such devices include spectrum analyzers, which require high spectral purity so a user can reliably distinguish authentic spectral components from quantization artifacts. While dithering is known to improve spectral purity, this property is only well-quantifiable for sinusoidal input signals through the SFDR metric. As a result, we restrict our scope to evaluation 
%of our post-quantization dithering via parametrized LUTs 
with highly-oversampled 
sinusoidal input signals. 
%quantized to low resolution.
This approach is a realistic use case while simplifying training of the model-based LUT  and, also, allowing for clear quantification of performance gains. Nevertheless, we are confident that the underlying advantage of the efficient digital estimation-dithering  technique is widely applicable to many input signal types not studied here.

%We decompose our model-driven parametrized LUT design into indexing, estimation, dithering, and quantization stages in 
The proposed design is given in
Section~\ref{sec:systemmodel}. The MSE-optimal high-resolution estimator is summarized in Section~\ref{sec:classicalest}. %, the evaluation of which is beyond the scope of this paper.
LUT dithering is elaborated in Section~\ref{sec:ditheringarch}. %including how to use dithering to control the trade-off between MSE and SFDR. Following this 
Masked indexing is studied in Sections~\ref{sec:bitmasksec} 
%, ~\ref{sec:hpisec}, 
% which support drastic reductions of LUT memory size detailed in 
and ~\ref{sec:memoryopt}.

\section{Preliminaries}
\subsection{Notation}
Table~\ref{tbl:notationtable} defines notation used throughout the paper.
%, where LC denotes lowercase and UC denotes uppercase. 
Not included in the table are constants, which can have arbitrary capitalization and subscripts but are explicitly stated to be constants (ex: $K$, $N$, $\rho$, sometimes $a,b$). Moreover, function definitions ($Q_{b}$), set definitions ($\mathcal{I}_{b}$), operators ($\mathbb{E}_{X}$), and estimators ($\hat{x}_{\mathrm{MMSE}}$) use their own independent subscript notation.
\begin{table}[h]
    \vspace{-4mm}
    \caption{Notation} \label{tbl:notationtable}
    \vspace{-4mm}
    \begin{center}
        \begin{tabular}{ |c|c| }
            \hline
            \textbf{Style} & \textbf{Interpretation} \\
            \hline
            Uppercase ($X$) & Random Variable (R.V.) \\ 
            Lowercase ($x$) & Realization of R.V. \\ 
            Bold Uppercase ($\mathbf{X}$) & Vector-Valued R.V. or Matrix \\
            Bold Lowercase ($\mathbf{x}$) & Vector \\ 
            Hat ($\hat{x}$) & Estimate \\
            %Tilde ($\,\widetilde{x}\,$) & Model/Approximation \\
            Calligraphic ($\mathcal{X}$) & Set or Transformation \\
            Subscript ($x_{n}$) & Time-Index ($x$ at time $n$) \\
            %UC with LC Subscript ($A_{z}$) & RV \\
            %UC with UC Subscript ($A_{L}$) & Constant
            %Capital Subscript ($X_{L}$) & Constant \\
            %{[Subscript]}
            Bracketed Subscript ($x_{[i]}$) & Vector-Index ($i$-th element of $\mathbf{x}$) \\
            \verb|Verbatim| & Stage of System Model \\
            \hline
        \end{tabular}
    \end{center}
    \vspace{-4mm}
\end{table}

% For convenience we define the rectangular function: 
% \begin{equation}
%     \mathsf{\Pi}_{a}(t) \triangleq 
%     \left\{\begin{matrix}
%     \frac{1}{a},& -\frac{a}{2} \leq t \leq \frac{a}{2}\\ 
%     0,& \textup{otherwise}
%     \end{matrix}\right.
% \end{equation}
%Note that for arbitrary $a,b$ we have $\mathsf{\Pi}_{(b-a)}(t-\frac{a+b}{2}) = \{ 1, t \in [a, b]; 0, t \notin [a,b]$.

We denote the probability distribution of a random variable $X$ %taking on the value $x$ as
with $p_{X}(x) = p(X=x)$. When written without the subscript, the random variable is implied 
%distribution is implied to be with respect to the corresponding R.V. 
(ex: $p(y|x) = p_{Y|X}(y | x) = p(Y=y| X=x)$). For a continuous R.V., $W \sim f_{W}(w)$ where $f_{W}(w)$ is its probability density function.
% When working with random variables, $\sim$ is used to denote that an R.V. (uppercase) takes on a certain distribution (ex: $W \sim \mathcal{N}(0, \sigma^{2})$) or that a realization of an R.V. (lowercase) is drawn from a certain distribution independently (ex: $\forall n, w_{n} \sim \mathcal{N}(0, \sigma^{2})$).

\subsection{Quantization}
%\section{Look-Up Table Architecture}
%Consider a 
The scalar quantization operation $Q(.)$ is defined on inputs $x \in \mathbb{R}$ with unique monotonically-increasing digital codebook in a vector $\bf{C}$ and unique monotonically-increasing analog partition levels in a threshold vector $\bf{T}$ such that: 
\begin{equation}\label{eq:quantizerdefn}
    Q(x) = C_{k},\;\; T_{k} < x < T_{k+1}
\end{equation}
where $k = \{1, \hdots, 2^{b}\}$ for a $b$-bit quantizer.
By convention the first partition value $T_{1} = -\infty$ and the last partition value $T_{2^{b}} = \infty$, ensuring $\mathrm{dom}(Q) = \mathbb{R}$.
%Behavior for the partition 
Edge cases are handled by modeling $Q(T_k)$ as a stochastic variable realizing $C_{k}$ or $C_{k-1}$ each with 
probability $1/2$, emulating a meta-stable comparator state. 
% For continuous input distributions the probability of this event is vanishing, allowing us to model the quantizer as functionally deterministic.
%Denote a finite-level with the subscript $Q_{b}(.)$, defined as having . 
%2^{b}$ entries in $\bf{C}$ and $2^{b}+1$ entries in $\bf{T}$.
A quantizer is \textit{uniform} if $C_{k+1}-C_{k} = T_{k+1}-T_{k} \triangleq \Delta$ for $k = \left\{2, \hdots, 2^{b}-1\right\}$. An infinite ($k \in \mathbb{Z}$) uniform quantizer can either be \textit{mid-tread} ($C_{k} = k\Delta, T_{k} = k\Delta - \Delta/2$) or \textit{mid-riser} ($C_{k} = k\Delta + \Delta/2, T_{k} = k\Delta$). In this paper we consider mid-riser quantizers since $T_{0} = 0$ ensures no dead-zone and, hence, the ability to represent arbitrarily low-amplitude signals. Furthermore, the $b$-bit uniform quantizers in this paper are normalized such that $\Delta = 2^{-b+1}$.

\subsection{Dithering}\label{sec:ditheringprelim}
% dithering
A \textit{dithered} quantizer forms the input as a sum of an analog input sample $x$ and an additive dither sample $w$. When used non-subtractively, the output is $y = Q(x+w)$ which maintains the same fixed-point resolution as the undithered quantization. 
% This is graphically illustrated in Fig.~\ref{fig:NSditherblockdiagram}. 
Dither can be an inherent property of the system, such as %when an analog-to-digital converter introduces 
Gaussian noise $W \sim \mathcal{N}(0, \sigma^{2})$ prior to quantization due to thermal effects.
% \begin{figure}[h]
%     \centering
%     \includegraphics[width=0.34\textwidth]{figures/LUTJRNL_NSDitherBD_V3.png}
%     \caption{Non-Subtractive Dither Block Diagram (Dashed Line Separates Analog/Digital)}
%     \label{fig:NSditherblockdiagram}
% \end{figure}
Alternatively, dither can be purposely added according to a particular distribution such as the popular uniform/rectangular dither $W \sim \mathrm{Uniform}(-\Delta/2, \Delta/2)$.
%\triangleq \mathsf{\Pi}_{\Delta}$
%or triangular ($W \sim \mathsf{\Pi}_{\Delta} \ast \mathsf{\Pi}_{\Delta}$, $\ast$ denotes convolution) dithers. 
Rectangular dither ensures that the quantizer is asymptotically unbiased (mean absolute error converges to 0 when averaging samples) with uncorrelated quantization error 
%, while the triangular dither also ensures constant (input-independent) mean square quantization error 
\cite{wannamaker1997thesis} \cite{lipshitzdithersurvey}.

To illustrate the advantage of dithering, consider a sinusoidal signal with weak additive white Gaussian noise quantized to 3-bit resolution.
% plot here comparison of three PSDs overlaid x, Q(x), Q(x+v)
\begin{figure}[h]
\centering
\includegraphics[width=0.48\textwidth]{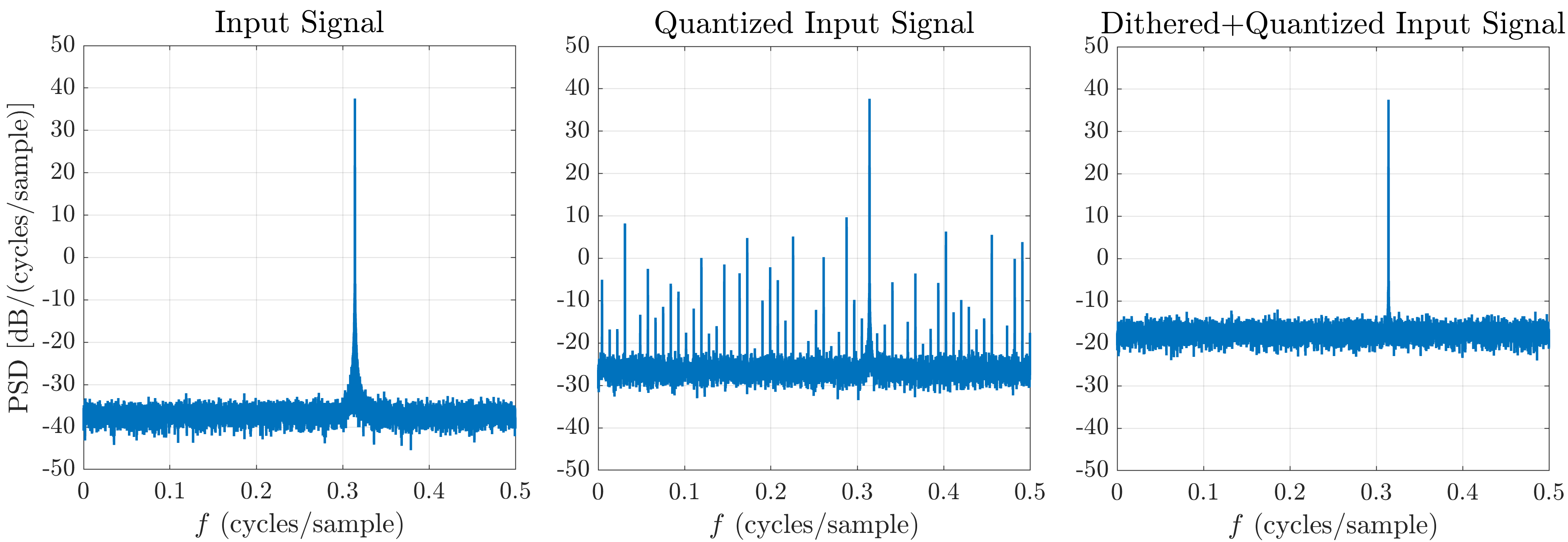}
\caption{Example PSD for Noisy Sinusoidal Input (Left) Quantized to 3-bit Resolution (Middle) and Quantized with Rectangular Dithering (Right)}
\label{fig:ditherpsdexample}
\end{figure}
Fig.~\ref{fig:ditherpsdexample} shows that without dithering the highly-correlated quantization error resulting from the low-resolution quantization generates harmonic ``spurs'' which alias throughout the output spectrum. 
%The result is a reduction in the SFDR of the system, as spectral components cannot be reliably distinguished as either authentic sinusoidal signals or simply quantization artifacts. 
By contrast, when using a uniform rectangular dither prior to quantization the output can maintain the same 3-bit resolution but with a significantly flatter spectral response, owing to the uncorrelated (spectrally white) quantization error. Consequently the SFDR of the system is improved despite an increase in the MSE (shown by the raised noise floor compared to the undithered quantization).
% describe here trade-off bw MSE and SFDR
Formally, the process of rectangular dithering raises the MSE of the resultant quantized signal by 3 dB relative to the undithered quantization (from $\Delta^{2}/12$ to $\Delta^{2}/6$). But as shown by Fig.~\ref{fig:ditherpsdexample}, this 3 dB MSE compromise can support a 20+ dB improvement in SFDR.
MSE and SFDR metrics are formally defined next.

\subsection{Figures of Merit}
Designing a digital post-correction scheme requires a metric or optimization objective for evaluation and comparison of any proposed methods. In practice, desired performance may be difficult to quantify (e.g., perceptual prominence of quantization artifacts) or conflict with alternative objectives (e.g., ease of implementation in a practical system). 
%In an effort to standardize these ideas we present three distinct metrics which we will use to evaluate our method.
Here, we describe three LUT evaluation metrics.

% MSE
Mean Square Error (MSE) is a classic and straightforward metric defined for a desired reference signal $\mathbf{x}$ and a test signal $\mathbf{\hat{x}}$ as:
\begin{equation}
    \mathrm{MSE\; [dB]} \triangleq 10 \log_{10}\left( \mathbb{E}\left[\left(\mathbf{x} - \mathbf{\hat{x}}\right)^{2}\right] \right)
\end{equation}

% SFDR
Spurious Free Dynamic Range (SFDR) is a frequency-domain metric intended to better represent the perceptual impact of our quantizer. This is important as the use of dithering strictly increases MSE, but can imbue the output signal with many desirable statistical properties such as uncorrelated and spectrally white error which are not captured by MSE alone. It is defined only for sinusoidal input signals with known fixed frequency $\tilde{f}$ as:
% \begin{equation}
%     \mathrm{SFDR\; [dBc]} \triangleq 10 \log_{10} \left( \frac{\left. \left|\mathcal{F}\left\{\mathbf{\hat{x}}\right\}\right|^{2}\right|_{f=f_{c}}} {\max_{f \neq f_{c}} \left|\mathcal{F}\left\{\mathbf{\hat{x}}\right\}\right|^{2}}\right)
% \end{equation}
% \begin{equation}\label{eq:sfdrdefn}
%     \mathrm{SFDR\; [dBc]} \triangleq 10 \log_{10} \left( \frac{ \left|\hat{X}(f_{c})\right|^{2}} {\max_{f \neq f_{c}} \left|\hat{X}(f)\right|^{2}}\right)
% \end{equation}
\begin{equation}\label{eq:sfdrdefn}
    \mathrm{SFDR\; [dBc]} \triangleq 10 \log_{10} \left( \frac{ \left|\hat{X}(\tilde{f}\,)\right|^{2}} {{\displaystyle \max_{f \notin [\tilde{f} - f_{\mathrm{o}}, \tilde{f} + f_{\mathrm{o}}]}} \left|\hat{X}(f)\right|^{2}}\right)
\end{equation}
where $\hat{X}(f) = \mathcal{F}\{\mathbf{\hat{x}}\} = \sum_{n} \hat{x}_{n} \exp(-j 2\pi f n)$ is used to estimate the Power Spectral Density (PSD) of the signal by computing its periodogram as $|\hat{X}(f)|^{2}$ and $f_{\mathrm{o}}$ is an offset term.
%intended to isolate components of the fundamental frequency from being included as spurious measurements. 
This offset term is necessary because SFDR is computed for a sequence of samples whose finite length will generate sidelobes due to windowing and generate spectral leakage due to non-integer period. 
%To avoid falsely characterizing either of these as quantization effects, we adopt $f_{\mathrm{o}} = 10^{-3}$ for our SFDR computations.
An example application where this criteria is critical is spectrum sensing or analysis, where the SFDR represents the maximum reliable dynamic range not containing quantization artifacts (``spurs'') which are detailed in Sec.~\ref{sec:ditheringprelim}.

%xHere we note that our conventional definition of SFDR in (\ref{eq:sfdrdefn}) uses the maximum value of the PSD at any non-input frequency. As a result, the strongest spur used for its calculation is typically a sidelobe of the desired signal's fundamental frequency. Some authors use a more robust definition of SFDR that involves computation of the maximum spur via Welch's method with windowing and a peak-finding algorithm to exclude any component of the fundamental, in which case the SFDR gain will be even greater than in the results shown in the following sections. Despite this alternative method producing numerically-superior results, it is excluded from this work due to the ad-hoc nature of designing an SFDR metric that relies on feature extraction rather than on the signal periodogram directly. With careful choice of window function, Welch's method parameters, and peak-finding criteria, results can easily be misrepresented. This problem is avoided entirely by using our much stricter SFDR definition.

% EVM
% In communication applications, Error Vector Magnitude (EVM) is a key metric due to its ability to quantify the likelihood of Bit Error Rate (BER) and hence the reliability of a communication link. It is defined in the root mean square (RMS) sense as:
% \begin{equation}
%     \mathrm{EVM\; [dB]} = 
% \end{equation}

% Memory
The memory size of a LUT is a key constraint on its practical implementations, as any FPGA or IC has an inherent limit (and associated cost) with the number of bits it must store. For a given LUT we denote the precision (resolution) of its stored entries in bits as $\rho$, and the number of entries it stores as $L$. We can express their impact on memory size as:
\begin{equation}\label{eq:memorysizedefn}
    \mathrm{Memory\; [bits]} \triangleq \rho \cdot L
\end{equation}
%This value is sometimes presented in bytes, where $\mathrm{bytes} = \lceil\mathrm{bits}/8\rceil$. 
Naturally, any proposed solution should include analysis of its required memory size to ensure it is feasible (and economical) to implement.

\section{System Model}\label{sec:systemmodel}
\subsection{Input Model}
% quantization problem x(kappa), w, y, etc.
We model at time-index $n$ the instantaneous input $x_{n}(\boldsymbol{\kappa}) + w_{n}$ % $w_{n}$ 
to the quantizer as a sum of two independent sources. A desired signal $x_{n}$ (parametrized by $K$ parameters $\boldsymbol{\kappa}$), and additive white noise (or dither) $w_{n}$ (which we assume to be iid drawn from stationary distribution $p(w)$). Parameters $\boldsymbol{\kappa}$ are, in general, random variables. The quantizer output is: 
\begin{equation}
    y_{n} = Q_{b}(x_{n}(\boldsymbol{\kappa}) + w_{n})
\end{equation}
%where $n$ is the quantized sample index.
For ease of analysis we assume without loss of generality that $C_{k} = k$  for $k = \{1, \hdots, 2^{b}\}$ (which can later be isomorphically transformed in the digital-domain to arbitrary $C_{k}$ as desired and, hence, without loss of generality). %This allows us to define
The set $\mathcal{I}_{b} \triangleq \left\{ 1, 2, \cdots, 2^{b} \right\}$ contains all possible quantization outputs at  resolution $b$. % We define
The index vector $\mathbf{y} \in \mathcal{I}_{b}^{N}$ contains the previous $N$ quantized samples as $\mathbf{y} = [y_{-N+1}, \cdots, y_{0}]^{T}$ such that $y_{[i]} = y_{-N+i}$.

\subsection{Look-Up Table Model}
% in/out relationship block diagram
\begin{figure}[h]
\centering
\includegraphics[width=0.48\textwidth]{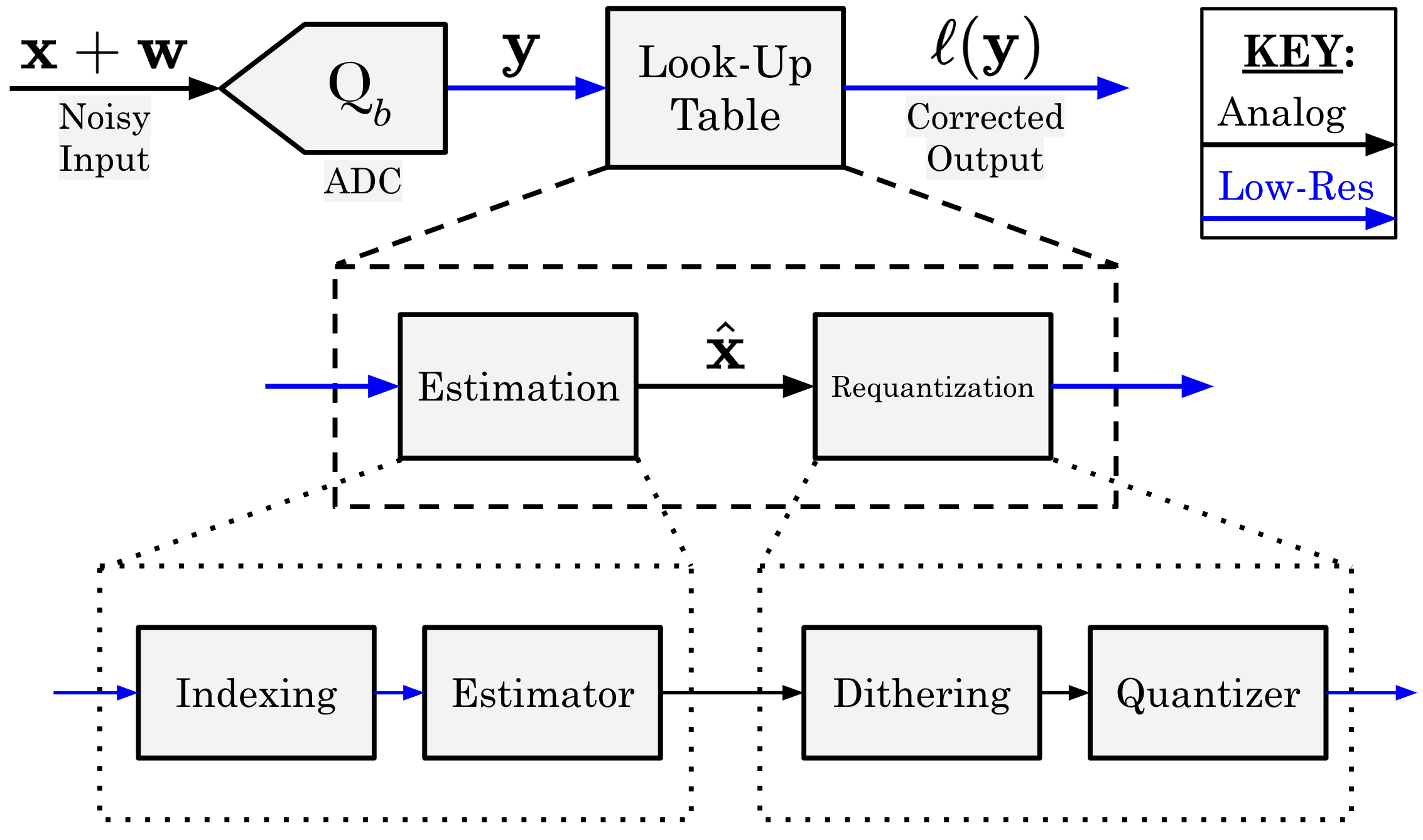}
\caption{Hierarchical Overview of Model for LUT Correction System}
\label{fig:LUTsystemmodel}
\end{figure}

% in/out l(yv)
Fig.~\ref{fig:LUTsystemmodel} describes the LUT correction approach at three levels of detail, outlined in this section. At the highest level, a LUT is simply a function 
$\ell : \mathcal{I}_{b}^{N} \rightarrow \mathcal{I}_{\rho}$ 
that (at each sample instant $n$) maps a quantized index sequence of length $N$ 
%$\mathbf{y} \in \mathbb{Z}^{N}$ 
to a quantized output value of arbitrary resolution $\rho$. 
%We are interested in LUTs that enforce $\rho=b$. 
%to ensure plug-and-play compatibility with any existing digital back-end. 
%$\mathbf{y} \rightarrow \ell(\mathbf{y})$
%with the same resolution as the input.
Designing the optimal LUT mapping is a difficult, intractable problem for sufficiently large $N$. 
%owing to its arbitrarily non-linear function definition with integer constraints on the input and output.
Optimality is subject to the individual design criteria of the system in which the ADC is embedded. For example, the optimal LUT is different for a spectrum sensing application, communication receiver, or data compression algorithm, since  each may seek to optimize for different metric. Further complicating this challenge is the desire to maintain the same quantization resolution at the LUT input and output ($\rho = b$) to ensure plug-and-play compatibility with existing throughput-limited digital systems. 
% Even quantifying optimality is difficult as it is subject to the individual design criteria of the system in which the ADC is embedded. For example, the optimal LUT is different for a spectral sensing application, communication receiver, or data compression algorithm, since  each may seek to optimize for different metric. Further complicating this challenge is the desire to maintain the same quantization resolution at the LUT input and output to ensure plug-and-play compatibility with existing throughput-limited digital systems. 
This problem has been conventionally addressed via numerical optimization of the LUT mapping using a data-driven calibration procedure. Such approach is unreliable, both because of the optimization step (non-convexity allows pre-mature convergence to local optima) and because of the use of training data (the quality of which limits the training accuracy). Any LUT trained in this way can only be proven optimal by brute-force evaluation of all possible LUTs with exponential search space. 
%\textcolor{red}{ Add your reference where you did this.}
Moreover, such methods are uninformative as the LUT is functionally treated as a black-box, with no way to diagnose under-performance. Lastly, integration of a numerically optimized LUTs with dithering/randomization stage is an open problem.
%as it requires re-training the entire table from scratch for each new signal model and for each different metric.

% problem decomposition into hi res est vs lo res dither requant 
To mitigate these shortcomings we decompose
%decompose the intractable LUT mapping problem 
the dithered LUT into two distinct components: an \verb|estimation| stage which seeks to recover the analog input signal with the highest fidelity, and a \verb|requantization| stage which seeks to represent that estimate in the same fixed-point resolution as the input signal with minimal loss of fidelity. We further decompose the \verb|estimation| stage into an \verb|indexing| scheme (transforming $\mathbf{y}$ into some alternative representation) followed by an \verb|estimator| (computing $\hat{x}(\mathbf{y})$ optimally). Moreover, we decompose the \verb|requantization| stage into a \verb|dithering| step (intended to condition the estimate for fixed-point representation while effectively trading-off the MSE increase with SFDR improvement) followed by a conventional \verb|quantizer|.

This structured decomposition of the LUT design problem has several advantages. First is reliability: each stage can be analytically-optimized, preventing sub-optimality due to numerical optimization. Second is that the LUT can be evaluated at each stage to diagnose reasons for under-performance, which allows the designer to intelligently select new design parameters. Third is the elimination of reliance on  training dataset (thus immunizing the system to such experimental errors) by adopting a model-driven training approach at each stage.

% estimation problem solution analysis yv
% In terms of the third hierarchical level (most detailed) in Fig.~\ref{fig:LUTsystemmodel}, our work studies two novel techniques for designing the \verb|indexing| stage in Sec.~\ref{sec:bitmasksec} and Sec.~\ref{sec:hpisec} as well as how to implement the \verb|dithering| stage in Sec.~\ref{sec:ditheringarch}. We briefly describe the MSE-optimal \textit{high-resolution} \verb|estimator| $\hat{x}_{\mathrm{MMSE}}(\mathbf{y})$ in Sec.~\ref{sec:classicalest}, the evaluation of which is beyond the scope of this paper. Optimization of the final \verb|quantizer| stage is also not studied in this work and assumed to be uniform.

\subsection{Example Input: Oversampled Tone}\label{sec:simsetup}
To illustrate the benefits of our proposed architecture we exclusively present simulated results using an input sinusoid of the form $x_{n}(A, F, \Phi) = A\cos(2\pi F n + \Phi)$. This is done to facilitate computation of the SFDR metric which is only well-defined for tone inputs. By oversampling the tone ($F < 0.5$) we emulate a wideband receiver preceding the ADC, which is an appropriate use case for the high-speed LUT-based correction we propose. The input parameters are fixed as $\boldsymbol{\kappa} = [A, F, \Phi]$ with $A = 1-\Delta/2 = 0.875$ and $F = \pi/10$ both assumed to be known a-priori to simplify training of the LUT by removing dependence on their prior distributions $p(\boldsymbol{\kappa})$. Note that the angular frequency $2\pi F = \pi^{2} / 5$ was intentionally chosen to be irrational and thus ensure ergodicity of the sequence $x_{n}$.

Each simulation generates $10^{5}$ samples using uniform mid-riser quantization with $b=3$-bit resolution ($\Delta = 0.25$). For convenience of notation, results use $Q(.) = Q_{3}(.)$ unless otherwise stated. The noise sequence is iid Gaussian $w_{n} \sim \mathcal{N}(0, \sigma^{2})$ with $\sigma/\Delta = 0.16$. To avoid falsely characterizing sidelobes of the fundamental frequency as spurs, we adopt $f_{\mathrm{o}} = 10^{-3}$ for our SFDR computed as per (\ref{eq:sfdrdefn}).

\section{Classical Estimation}\label{sec:classicalest}
The Minimum Mean Square Error (MMSE) %\verb|estimator| 
estimator is   \cite{kay1993fundamentals}:
\begin{equation}
    \hat{x}_{0, \text{MMSE}}(\mathbf{y}) = 
    \frac{\int_{-\infty}^{\infty}x_{0} \cdot p(x_{0}) \cdot p(\mathbf{y} | x_{0}) dx_{0}}{\int_{-\infty}^{\infty} p(x_{0}) \cdot p(\mathbf{y} | x_{0}) dx_{0}}\label{eq:mmseformulation}
\end{equation}
where (in our case) we have
\begin{equation}\label{eq:pyvcondx0_final}
    p(\mathbf{y} | x_{0}) = \idotsint_{\mathbb{R}^{K}} p(\boldsymbol{\kappa} | x_{0}) \cdot \left(\prod_{n=-N+1}^{0} p(y_{n} | \boldsymbol{\kappa})\right) d\boldsymbol{\kappa}
\end{equation}
with
\begin{equation}\label{eq:pyncondku}
    p(y_{n} | \boldsymbol{\kappa}) \\
    = \int_{T_{y_{n}} - x_{n}(\boldsymbol{\kappa})}^{T_{y_{n}+1} - x_{n}(\boldsymbol{\kappa})} p(w) dw
\end{equation}

Note that for Gaussian noise/dither $W \sim \mathcal{N}(0, \sigma^{2})$ we have:
\begin{equation}
    \int_{a}^{b} p(w) dw = \frac{1}{2}\left[\text{erf}\left(\frac{b}{\sigma \sqrt{2}}\right) - \text{erf}\left(\frac{a}{\sigma \sqrt{2}}\right) \right]
\end{equation}
for arbitrary $a,b$.
%In all equations, $p(x_{0})$ denotes 
The prior distribution of the instantaneous input signal sample is:
%which depends on  $\boldsymbol{\kappa}$  and, hence,:
\begin{equation}\label{eq:priorx0kappa}
    p(x_{0}) = \idotsint_{\mathbb{R}^{K}} p(\boldsymbol{\kappa})\cdot p(x_{0} | \boldsymbol{\kappa}) \,d\boldsymbol{\kappa}
\end{equation}

For the tone signal in Sec.~\ref{sec:simsetup}, we have:
%$\forall x_{0} \in (-A, A)$:
\begin{align}
    p(x_{0}) &= \frac{1}{\pi\sqrt{A^{2} - x_{0}^{2}}} \cdot \mathbbm{1}_{x_{0} \in (-A,A)}\\
    p(\boldsymbol{\kappa}|x_{0}) &= p(a,f|x_{0}) \cdot p(\phi | a,f,x_{0})\nonumber\\
    &= \frac{1}{2} \sum_{m=0}^{1} \delta \left(\phi + (-1)^{m} \arccos\left(\frac{x_{0}}{A}\right)\right) \cdot \mathbbm{1}_{x_{0} \in (-A,A)}
\end{align}
where $\delta(.)$ denotes the dirac delta function and $\mathbbm{1}_{x_{0} \in (-A,A)}$ is an indicator function equal to 1 if $x_{0} \in (-A,A)$ and 0 otherwise. Notably these distributions do not depend on the tone frequency $F$, which holds only under the assumption of ergodicity $(1/f \notin \mathbb{Z})$ made in Sec.~\ref{sec:simsetup}.
%and large observation interval relative to the period of the signal.

\section{Dithering Architecture}\label{sec:ditheringarch}
Dithering is an inherently stochastic process. To implement it in a digital LUT correction architecture as per the \verb|dithering| stage in Fig.~\ref{fig:LUTsystemmodel} requires careful treatment of this stochastic behavior to preserve its desirable statistical properties. To this end we propose three different dithering architectures, the properties of which are qualitatively summarized in Table~{\ref{tbl:ditheringcomparison}} for a LUT with $L$ total entries.

\begin{table}[h]
    \vspace{-4mm}
    \caption{Proposed LUT Dithering Architectures} \label{tbl:ditheringcomparison}
    \vspace{-4mm}
    \begin{center}
        \begin{tabular}{ |c|c|c| }
            \hline
            \textbf{Method} & \textbf{Description (per Indexing Sequence)} & \textbf{Memory Cost} \\
            \hline
            \multirow{2}{*}{Intra-Table} & Hard-Code One Table ($\Xi = 1$) & \multirow{2}{*}{$b \cdot L$} \\
            & with Single Dither Realization & \\
            \hline
            \multirow{2}{*}{Inter-Table} & Multiplex $\Xi$ Tables of & \multirow{2}{*}{$\Xi \cdot b \cdot L$} \\
            & Independent Dither Realizations & \\
            \hline 
            \multirow{2}{*}{Post-Table} & Index High-Precision Estimate, & \multirow{2}{*}{$\rho \cdot L$} \\
            & Dither Stochastically and Requantize & \\
            \hline
        \end{tabular}
    \end{center}
    \vspace{-4mm}
\end{table}

% define parametric dithering (alpha)
Choice of the optimal distribution $p(v)$ for the dither random variable is non-trivial and the subject of extensive literature beyond the scope of this paper. For this work we adopt the parametric dither distribution proposed in \cite{klimesh_quantization_1999} \cite{klimesh_optimal_2000}, and studied in further detail in \cite{kasher2024dcc_direct} \cite{kasher2024dcc_arxiv} as:
\begin{equation}\label{eq:ditherdist}
    p(v) 
    %= \mathsf{\Pi}_{\alpha\Delta}(v) 
    = \left\{\begin{matrix}
    \frac{1}{\alpha\Delta},& -\frac{\alpha\Delta}{2} \leq v \leq \frac{\alpha\Delta}{2}\\ 
    0,& |v| > \frac{\alpha\Delta}{2}
    \end{matrix}\right.
\end{equation}
where $\alpha \in [0,1]$ and $\Delta = 2^{-b+1}$. Intuitively $\alpha$ represents the peak amplitude (bounded by the quantization interval $\Delta$) of the dither random variable which maintains a rectangular (uniform) distribution. Functionally, $\alpha$ represents the trade-off between MSE and SFDR, both of which increase with $\alpha$.

\subsection{Intra-Table Dithering}
\begin{figure}[h]
    \centering
    \includegraphics[width=0.48\textwidth]{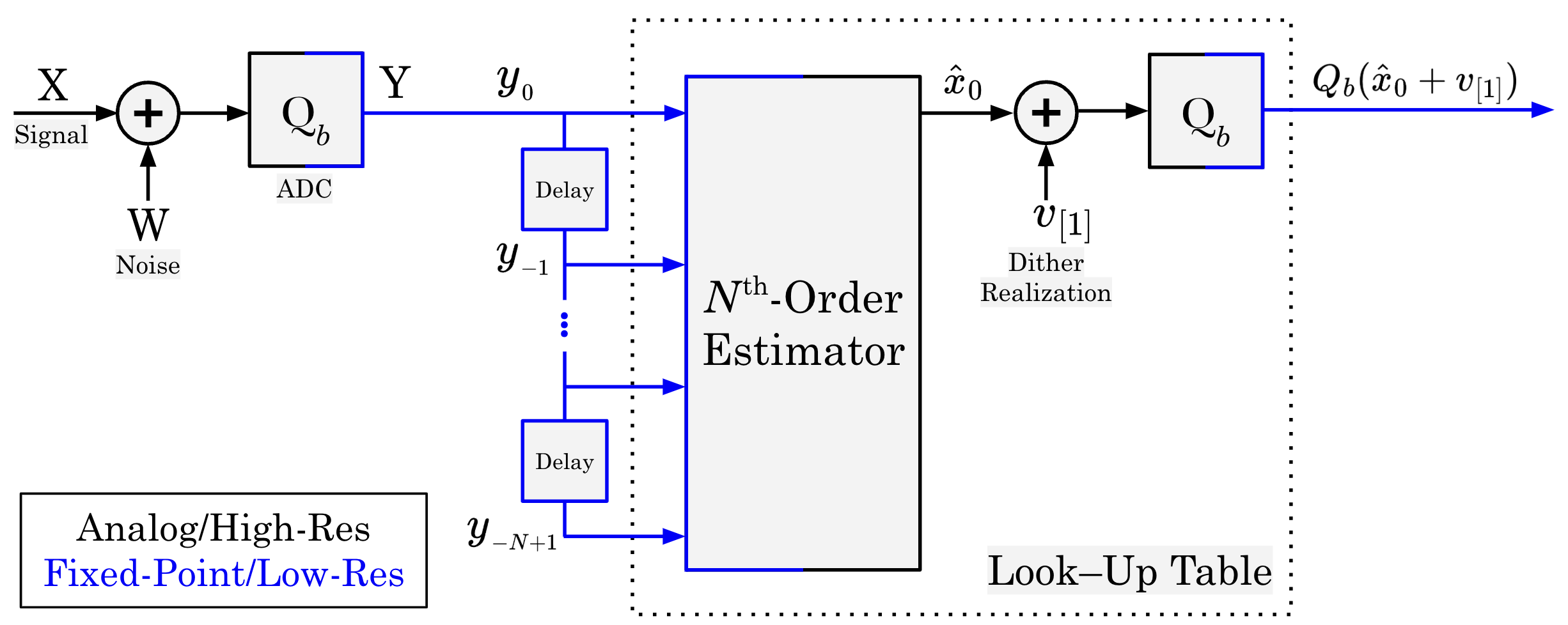}
    \caption{Intra-Table Dithering Architecture}
    \label{fig:intratableditherarch}
\end{figure}
The simplest method of dithering is to generate a single realization of the dither random variable for each LUT entry and hard-code all entries directly into one table, as shown in Fig.~\ref{fig:intratableditherarch}. 
This is done by taking the high-precision estimator output and adding to it a high-precision dither value (single realization) before requantizing it to the fixed-point output precision and storing the result as a single LUT entry for direct-indexing.
%This is done by generating a single fixed dither realization $v_{[1]}$ per indexing sequence $\mathbf{y}$ and storing the fixed-point output of each LUT index as $Q_{b}(\hat{x}_{0}(\mathbf{y}) + v_{[1]}(\mathbf{y}))$.
%Mathematically the LUT output for one particular indexing sequence is $Q_{b}(\hat{x}_{0} + v_{[1]})$, where $v_{[1]}$ denotes one fixed realization of $V$.
The resultant LUT output in the notation of Fig.~\ref{fig:LUTsystemmodel} is of the form:
\begin{equation}
    \ell(\mathbf{y}) = Q_{b}(\hat{x}_{0}(\mathbf{y}) + v_{[1]}(\mathbf{y}))
\end{equation}
where $v_{[1]}$ denotes one 
%fixed 
realization of $V$ per indexing sequence $\mathbf{y}$.
We denote this method as Intra-Table Dithering, and is a special-case of the Inter-Table Dithering introduced next  with $\Xi = 1$. The advantage of this technique is its minimal memory requirement and simplified implementation, but at the expense of reduced dither effectiveness.

\subsection{Inter-Table Dithering}
\begin{figure}[h]
    \centering
    \includegraphics[width=0.48\textwidth]{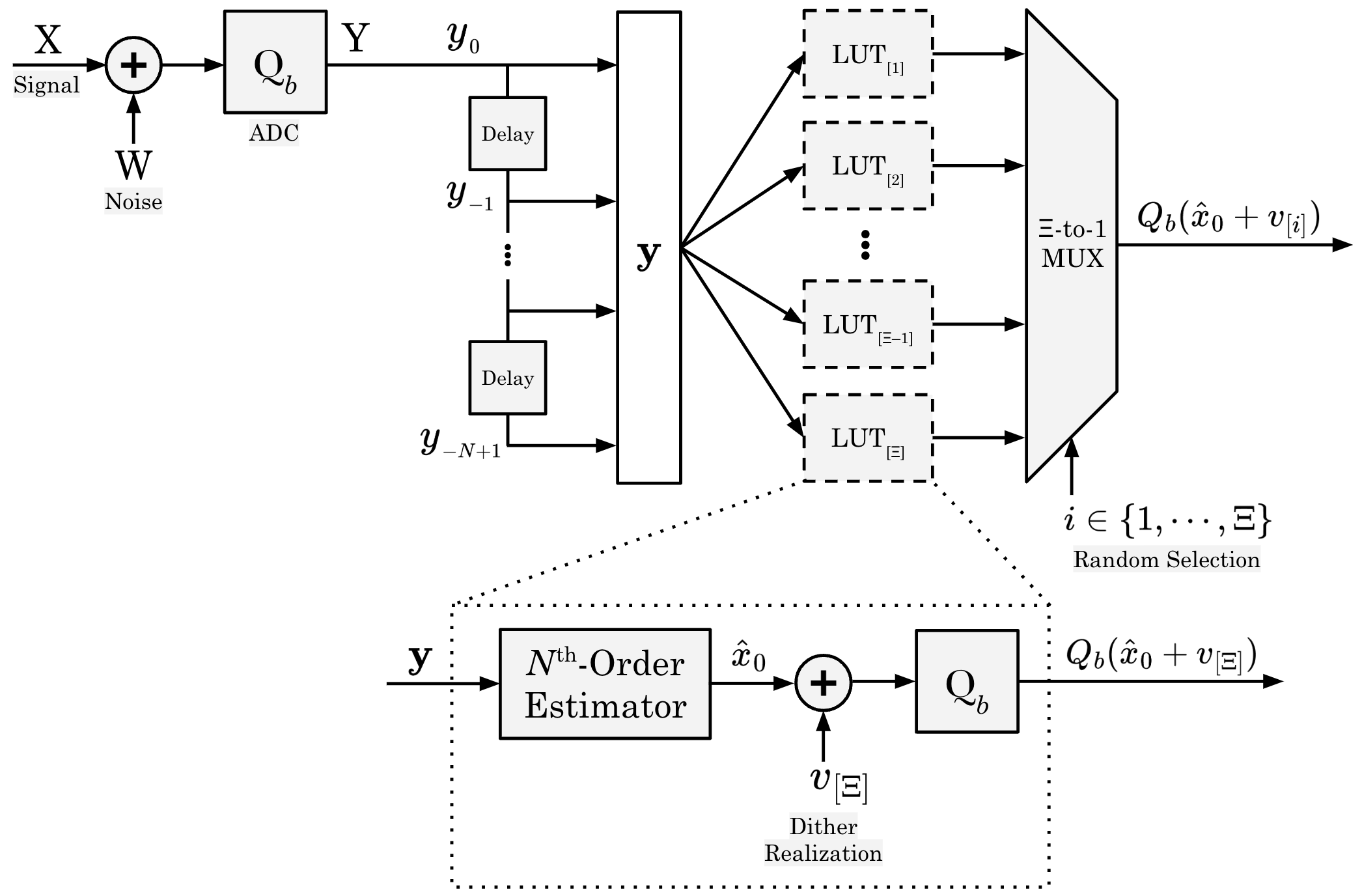}
    \caption{Inter-Table Dithering Architecture}
    \label{fig:intertableditherarch}
\end{figure}
Rather than generating a single LUT for direct indexing, consider the generation of $\Xi$  look-up tables. Each time a LUT entry is indexed, one of the $\Xi$ tables is randomly selected to produce the instantaneous output as illustrated in Fig.~\ref{fig:intertableditherarch}. 
In this scheme the resultant stochastic LUT mapping is of the form:
\begin{align}
    \ell(\mathbf{y}) &= Q_{b}(\hat{x}_{0}(\mathbf{y}) + v_{[i]}(\mathbf{y}))\\
    i &\sim \mathrm{Uniform}\{1, \Xi\}\nonumber
\end{align}
where $\mathrm{Uniform}\{1, \Xi\}$ is the discrete uniform distribution taking  values in $\{1, \cdots, \Xi\}$ 
%each with probability $1/\Xi$ 
and $v_{[i]}(\mathbf{y})$ denotes a realization of $V$ corresponding to an indexing sequence $\mathbf{y}$.
%Each LUT being multiplexed stores fixed-point entries $Q_{b}(\hat{x}_{0}(\mathbf{y}) + v_{[i]}(\mathbf{y}))$, where $i = \{1, \cdots, \Xi\}$ is the random selection performed by the multiplexer each with probability $1/\Xi$ and $v_{[i]}$ denotes one fixed realization of $V$ for a given indexing sequence $\mathbf{y}$.
%The set of LUT entries for a given index is produced by computing the high-precision estimate $\hat{x}_{\mathrm{MMSE}}$ for that indexing sequence, generating $\Xi$ realizations of the dither random variable, and for each such realization adding the dither value to the estimate before requantizing to the fixed-point output precision and storing it as an independent LUT entry. 

In this way, 
%$\Xi$ dither values are realized per indexing sequence, producing $\Xi$ fixed-point entries distributed over $\Xi$ tables. 
%The 
the stochastic dither process is 
%consequently 
replaced by a stochastic \textit{indexing} process, for which each result is hard-coded allowing efficient real-time access with no  dithering  nor requantization at run-time. 
%generation, addition, or . 
%Mathematically the LUT output for one particular indexing sequence is $Q_{b}(\hat{x}_{0} + v_{[i]})$, where $i = \{1, \cdots, \Xi\}$ each with probability $1/\Xi$ and $v_{[i]}$ denotes one fixed realization of $V$. 
This approach is termed Inter-Table Dithering, with the efficiency of the dithering and total memory requirement necessarily a function on the number of tables $\Xi$.
%used while the total memory requirement to store the full set of LUTs is directly proportional to $\Xi$.

\subsection{Post-Table Dithering}\label{sec:posttabledithering}
\begin{figure}[h]
    \centering
    \includegraphics[width=0.48\textwidth]{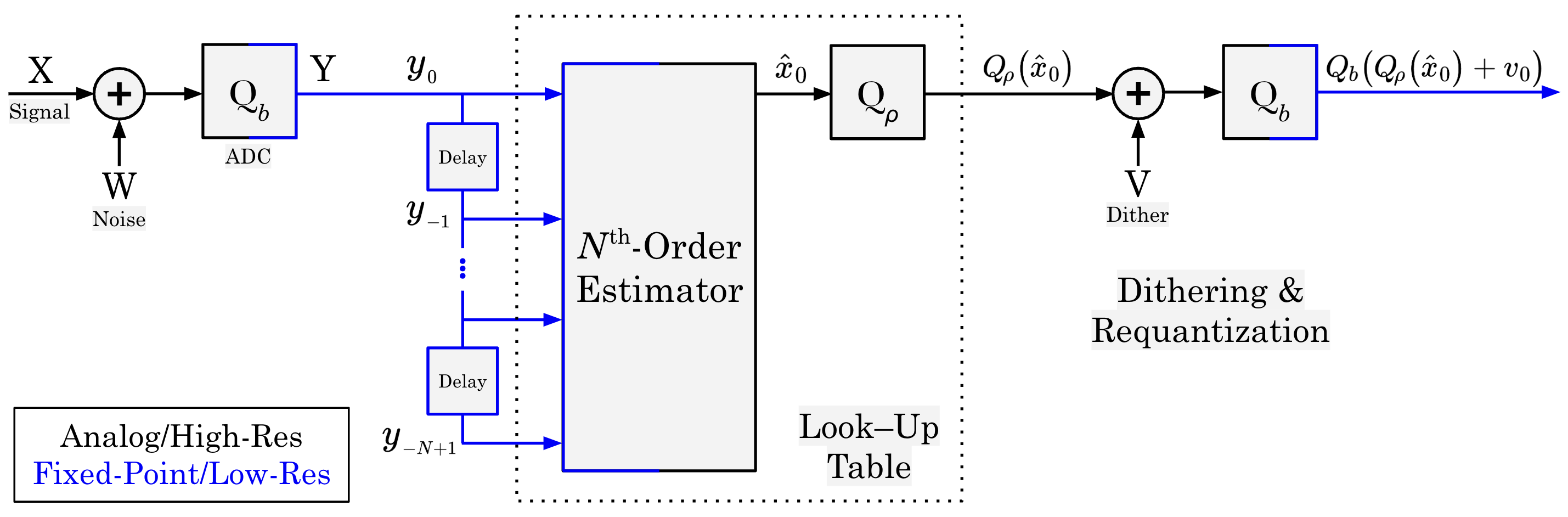}
    \caption{Post-Table Dithering Architecture}
    \label{fig:posttableditherarch}
\end{figure}
Finally, 
%most authentic (true-to-form with the reference analog method) dithering architecture is to store
a high-precision estimate $\hat{x}_{0}(\mathbf{y})$ 
can be stored 
in a single LUT with fixed-point precision $\rho$. 
%When indexing the LUT this estimate is directly accessed, output in high-precision, and then added digitally with 
A new realization of the dither is added to the each LUT output, before being requantized to a lower  resolution. This process is illustrated in Fig.~\ref{fig:posttableditherarch} and produces output 
\begin{equation}
    \ell(\mathbf{y}) = Q_{b}(Q_{\rho}(\hat{x}_{0}(\mathbf{y})) + v_{0})
\end{equation}
where $v_{0}$ is
%is a new instantaneous random
a dither realization independent of $\mathbf{y}$  also of resolution  $\rho$. 

All of this must take place in real-time, increasing the computational burden and overhead.
%to the LUT system. 
Moreover, high-precision estimation significantly   increases the storage requirements. Notably, the post-table dithering stage requires data transfer with $\rho$ bits of resolution at full-rate (real-time). 
%Since $\rho \gg b$, when the digital back-end is rate-limited then the  dithering is not feasible. 
This is the only architecture that maintains all desired statistical properties of the dither signal. 
%Mathematically the LUT output for one particular indexing sequence is $Q_{b}(Q_{\rho}(\hat{x}_{0}) + v)$, where $v$ denotes a new random realization of $V$. 
It is denoted Post-Table Dithering, and its performance is equivalent to that of Inter-Table Dithering in the limit as $\Xi \rightarrow \infty$ since generating each dither realization independently at run-time is equivalent to generating infinitely many independent dither realizations and storing them ahead of time.

Note that since this is the only method where the dither signal is generated as part of the fixed-point signal chain, the choice of distribution $p(v)$ is necessarily discrete rather than the continuous one proposed in (\ref{eq:ditherdist}). Nevertheless, we can generate a discrete dither equivalent to the continuous one~ \cite{kollar_digital_2006}.
% In fact, \cite{kollar_digital_2006} proves that the result after quantization will be indistinguishable to the one when an equivalent digital dither with resolution $\rho$ is used.

\subsection{Comparison of Dither Strategies}
\begin{figure}[h]
    \centering
    \includegraphics[width=0.48\textwidth]{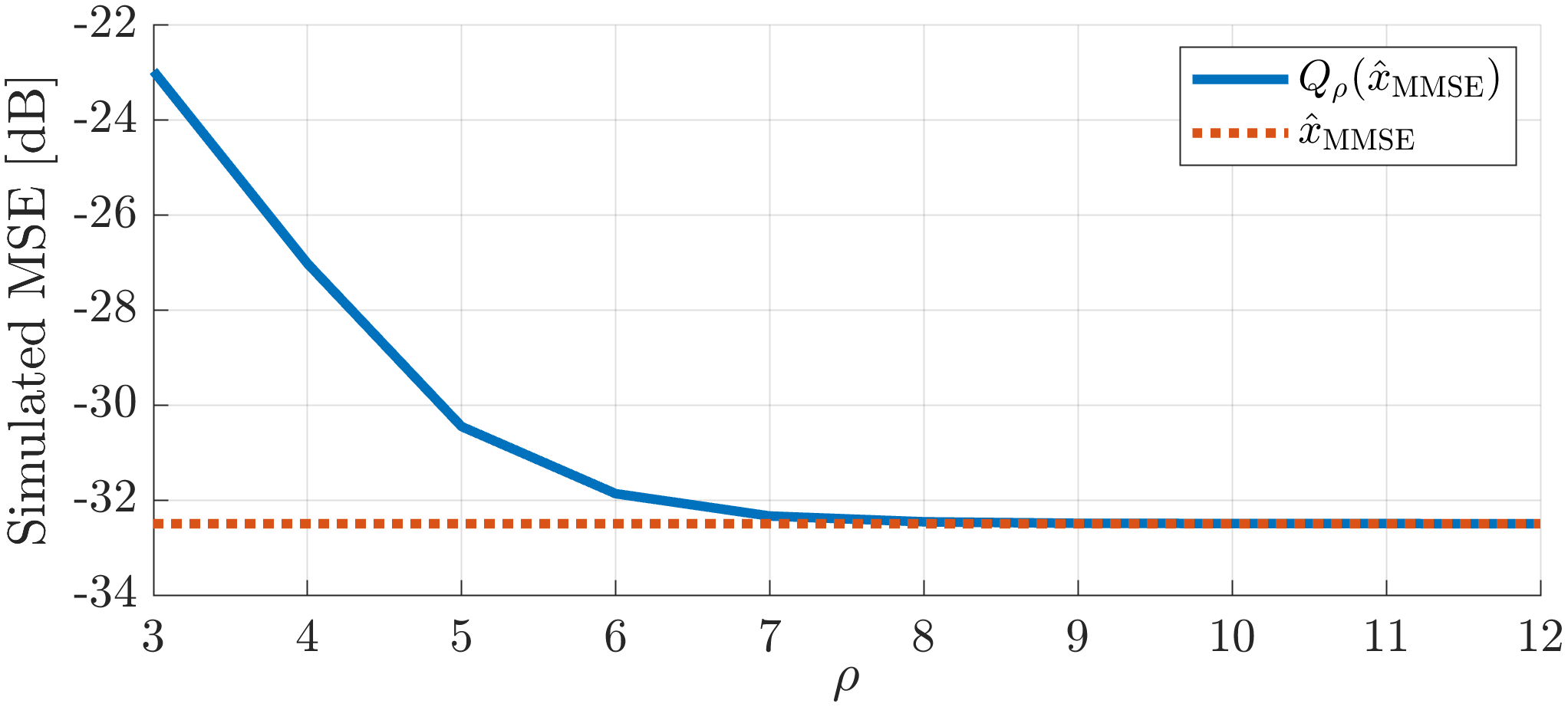}
    \caption{Effect of Requantization with Varying Resolution $\rho$ on Estimate of Quantized Tone with $N=10$}
    \label{fig:MSEoverQRho}
\end{figure}

%Simulation setup parameters are the same as those described in Sec.~\ref{sec:simsetup}. 
% In order to functionally implement post-LUT dithering, 
% we must first select an appropriate value of $\rho$ to store our high-precision input estimate. 
% Due to estimation error, 
% We study the dither strategies by evaluating the  Effective Number of Bits (ENOB) of the LUT output.
% The effective Number of Bits (ENOB) of the LUT output, a classic ADC metric, is a function 
% the estimation error and hence  on both $N$ and the LUT resolution  $\rho.$ And, as an example,  Fig.~\ref{fig:MSEoverQRho} illustrates that, for a quantized $b=3$-bit tone estimated using $N=10$, there is little 
The estimation error is a function of  both index size $N$ and the LUT resolution  $\rho.$    Fig.~\ref{fig:MSEoverQRho} illustrates that, for a quantized $b=3$-bit tone estimated using $N=10$, in this example there is little benefit to storing more than $\rho=8$-bit precision. 
%It is desirable to use the minimum allowable value of $\rho$ since memory size of the LUT is directly proportional to $\rho$.
% figure showing requantization impact (rho=8)

\begin{figure}[h]
    \centering
    \includegraphics[width=0.48\textwidth]{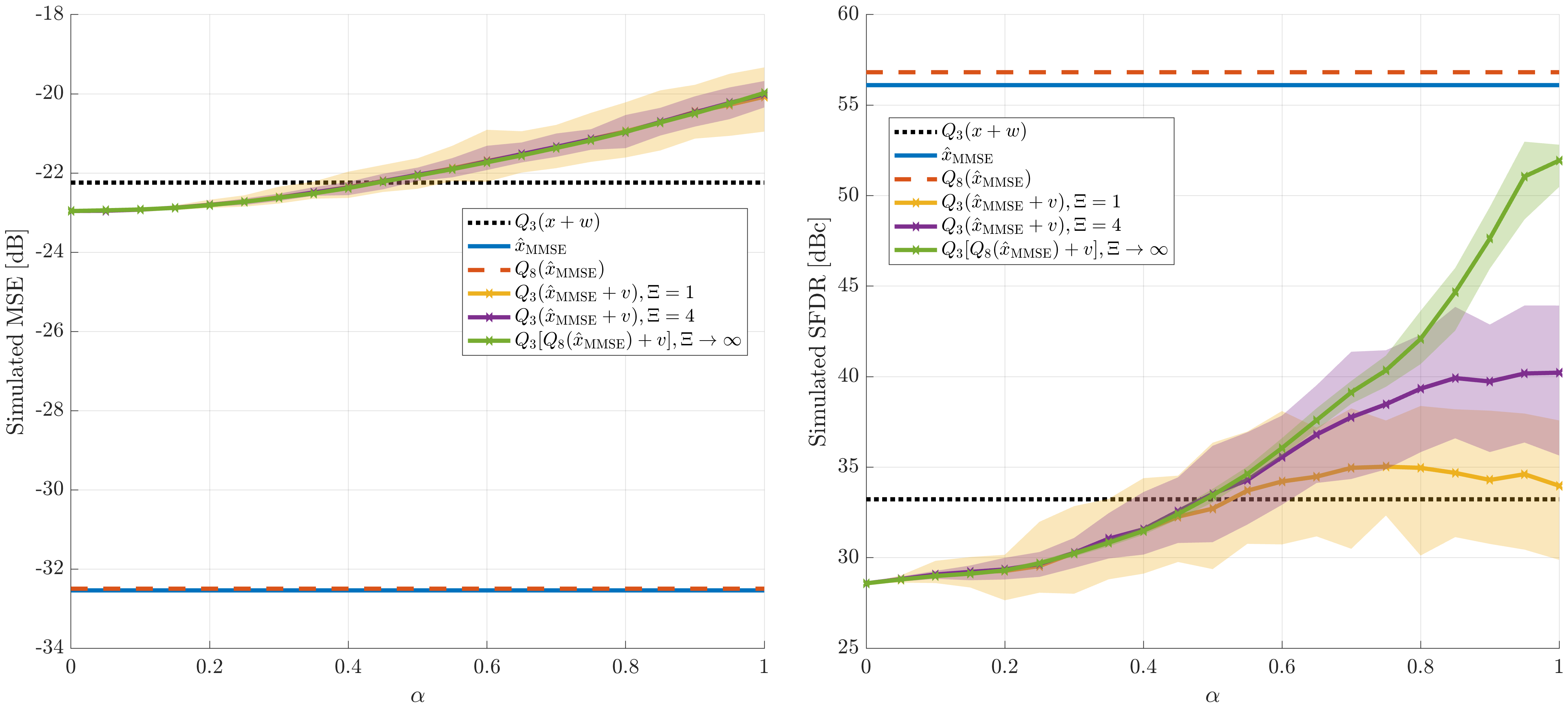}
    \caption{Effect of Dither Amplitude $\alpha$ on Requantized Estimate of Simulated Quantized Tone with $N=10$ for Varying Dithering Architectures (Intra-Table, Inter-Table, Post-Table). Each $\alpha$ Value Simulates 100 Independent Trials, with Average Performance (Solid Line) and Max/Min Performance (Shaded Region) Both Shown.}
    \label{fig:DitheringMethodEvaluation}
\end{figure}

%\subsection{Dithering Evaluation}
All three dithering techniques are directly compared as a function of $\alpha$ using $\Xi = 1$ (intra-table dithering), $\Xi = 4$ (inter-table dithering), and $\Xi \rightarrow \infty$ (post-table dithering) for a simulated quantized tone in Fig.~\ref{fig:DitheringMethodEvaluation}. ( See Sec.~\ref{sec:simsetup}. for the simulation setup.)

\noindent Several key insights are revealed by this result:
\begin{itemize}
    \item Dithering always worsens MSE (by up to 3 dB) but improves SFDR (by up to 19+ dBc), making it ideal for applications where spectral purity and dynamic range are desirable over strict error metrics. 
    % \item The performance  is highly dependent on the actual realizations of the dither, with significant variance across each of the 100 independent trials simulated for each $\alpha$ value. This is shown by the maximum SFDR deviation of up to 5 dBc from average performance. Increasing $\Xi$ appears to reduce this variance. 
    \item The performance  is highly dependent on the actual realizations of the dither, exhibiting a significant variance which grows with $\alpha$. The maximum SFDR deviation  from average is up to 5 dBc. Increasing $\Xi$ appears to reduce this variance. 
% Choice of optimal dither realizations for a given $\Xi$ to reliably achieve the top-end of the SFDR distribution is a topic for future work.
    \item The SFDR-optimal dither amplitude is not always equal to $\alpha=1$, sometimes peaking at values around $\alpha \in [0.8, 0.9]$ for average and/or best-case performance of the architectures tested. Dithering with $\alpha < 0.3$ never improves SFDR while $\alpha < 0.5$ typically does not either.
    \item Increasing $\Xi$ substantially improves both the average and best-case SFDR performance.
% of the LUT for SFDR optimization,
Nevertheles, $\Xi$ has little effect on MSE beyond reducing its variance.
    \item Post-table dithering is always the preferred architecture, 
%if the design constraints allow its implementation,
as it has the highest average and best-case SFDR improvement with the lowest variance. It is able to achieve as much as \textit{19+ dBc SFDR improvement} relative to the input samples. 
    \item The maximum achieved SFDR by post-table dithering after requantization is almost exactly 3 dBc lower than that of the high-resolution estimate, a fact which is neatly accounted for by the 3 dB analytical MSE increase uniformly raising the noise floor of the output PSD. This suggests that post-table dithering may be close to optimal in the sense that it achieves close to  maximum possible SFDR improvement at the \verb|requantization| stage (for the dither distribution in (\ref{eq:ditherdist})).
\end{itemize}

\section{Bit-Masking}\label{sec:bitmasksec}
 Bit-masking  \cite{lundin_criterion_2004} \cite{lundin_optimal_2005} aims at indexing the LUT using a subset of bits taken from the dyadic expansion of the input sequence $\mathbf{y}.$ %The size of the masked index subset is $\beta \leq bN.$
 %at the \verb|indexing| stage in Fig.~\ref{fig:LUTsystemmodel}.
 Here, we study how to implement the bit-masking following Fig.~\ref{fig:LUTsystemmodel}.

\begin{figure}[h]
    \centering
    \includegraphics[width=0.48\textwidth]{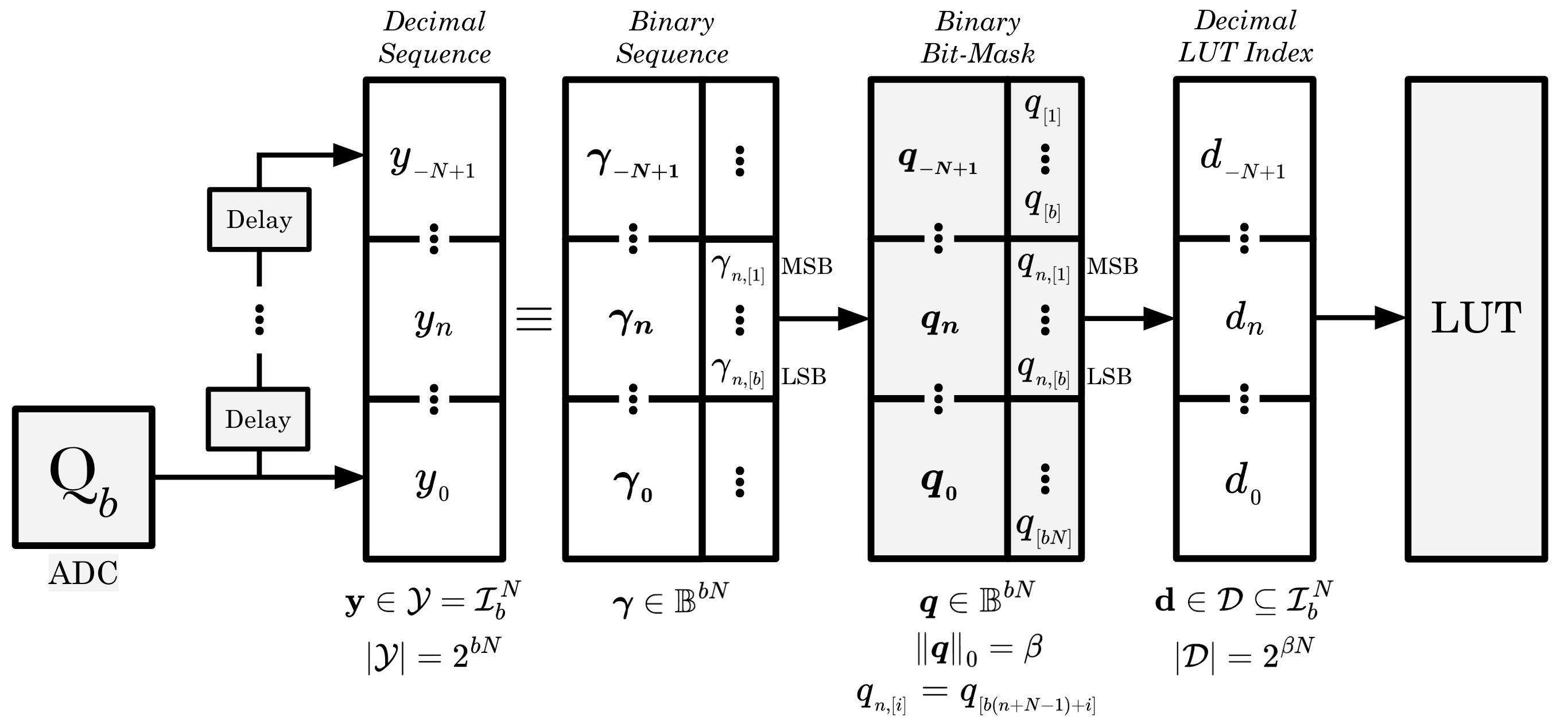}
    \caption{Block Diagram for Indexing a LUT with Bit-Masking}
    \label{fig:BitMaskBD}
\end{figure}

\subsection{Preliminaries}
Consider the binary set $\mathbb{B} = \left\{ 0, 1 \right\}$. Let $\mathbf{y} \in \mathcal{Y} = \mathcal{I}_{b}^{N}$, whose elements
%and ternary set $\mathbb{B_{\chi}} = \left\{ 0, 1, \chi \right\}$ with binary values and ``don't care'' value $\chi$ which can take on arbitrary binary value. 
%
%For a given value
$y_{n} \in \mathcal{I}_{b}$ have a dyadic expansion:
\begin{equation}\label{eq:dyadicexpansionyn}
    y_{n} = 1 + \sum_{i=1}^{b} \gamma_{n,[i]} \cdot 2^{b-i}
\end{equation}
where $\boldsymbol{\gamma_{n}} \in \mathbb{B}^{b}$ is ordered from the Most Significant Bit (MSB) $\gamma_{n,[1]}$ to the Least Significant Bit (LSB) $\gamma_{n,[b]}$.

Define a bit-mask as the binary vector $\boldsymbol{q} \in \mathbb{B}^{bN}$ where $\beta \triangleq \left\| \boldsymbol{q} \right\|_{0} \leq bN$ is the  size of the bit-mask.
The bit-mask indices $\boldsymbol{q_{n}} \in \mathbb{B}^{b}$ satisfy $q_{n,[i]} = q_{[b(n + N - 1)+i]}$. The bit-mask is selected by the designer and hence always known a-priori.
%, allowing us to condition on it in subsequent expressions.
Define $\mathbf{d}$ as the decimal index vector used as the input to the LUT, created by applying bit-mask $\boldsymbol{q}$ to $\mathbf{y}$.
The decimal elements of $\mathbf{d}$ are:
\begin{equation}\label{eq:decimalexpansiondn}
    d_{n} = 1 + \sum_{i=1}^{b} \gamma_{n,[i]} \cdot q_{n,[i]} \cdot 2^{b-i}
\end{equation}

A system-level block diagram of the bit-masking operation using this notation is shown in Fig.~\ref{fig:BitMaskBD}. With this framework we can can express the MMSE estimate as:
%that the LUT is intended to implement as:
\begin{equation}\label{eq:mmseformulationdv}
    \hat{x}_{0, \text{MMSE}}(\mathbf{d}|\boldsymbol{q}) = \frac{\int_{-\infty}^{\infty}x_{0} \cdot p(x_{0}) \cdot p(\mathbf{d} | x_{0}, \boldsymbol{q}) dx_{0}}{\int_{-\infty}^{\infty} p(x_{0}) \cdot p(\mathbf{d} | x_{0}, \boldsymbol{q}) dx_{0}}\end{equation}
%\begin{align}
%    d_{n}(\boldsymbol{q_{n}}, \boldsymbol{\gamma_{n}}) &= \sum_{i=1}^{b} \gamma_{n,[i]} \cdot q_{n,[i]} 
%\cdot 2^{b-i}\nonumber\\
%    d_{n} &= 
%\end{align}
%We can define the vector containing these scalars over time as $\mathbf{d}$. 
%This allows us to express:
where $p(\mathbf{d} | x_{0}, \boldsymbol{q})=\\$
% re-write p(yv | x0) for d
\begin{multline}\label{eq:pdcondx0q}
    = \idotsint_{\mathbb{R}^{K}} p(\boldsymbol{\kappa} | x_{0}) \cdot \left(\prod_{n=-N+1}^{0} p(d_{n} | \boldsymbol{\kappa}, \boldsymbol{q_{n}})\right) d\boldsymbol{\kappa}
\end{multline}

Now with slight abuse of notation we define $\boldsymbol{\gamma_n}(y_{n}) : \mathcal{I}_{b} \rightarrow \mathbb{B}^{b}$ to be the dyadic expansion of scalar $y_{n}$ as per (\ref{eq:dyadicexpansionyn}) and $d_{n}(\boldsymbol{\gamma_{n}}, \boldsymbol{q_n}): (\mathbb{B}^{b},\mathbb{B}^{b}) \rightarrow \mathcal{I}_{b}$ to be the decimal representation of the binary vector $\boldsymbol{\gamma_{n}}$ bit-masked by $\boldsymbol{q_{n}}$ as per (\ref{eq:decimalexpansiondn}). Then:
% express p(yn | K, U) for d
\begin{equation}
    p(d_{n} | \boldsymbol{\kappa}, \boldsymbol{q_{n}}) = \sum_{y_{n}\, :\, d_{n}(\boldsymbol{\gamma_{n}}(y_{n}), \boldsymbol{q_{n}}) = d_{n}} p(y_{n} | \boldsymbol{\kappa})
\end{equation}
where $p(y_{n} | \boldsymbol{\kappa})$ is given by (\ref{eq:pyncondku}).
Define $\mathcal{D}$ as the set of all possible bit-masked decimal indexing sequences. $\mathcal{D} \subseteq \mathcal{Y}$ is a function of the bit-mask $\boldsymbol{q}$, but for notational convenience we omit this dependency. 
%it is implied that $\mathcal{D} = \mathcal{D}(\boldsymbol{q})$. 
Further, define the conditional support set $\mathcal{D}'(x_{0}) \triangleq 
%\mathrm{supp}(\mathcal{D}) = 
\{\mathbf{d} \in \mathcal{D} \;|\; p(\mathbf{d} | x_{0}, \boldsymbol{q}) \neq 0\}$. 
% Note that for dither signals with infinite support such as $W \sim \mathcal{N}(0, \sigma^{2})$ we have $(\mathrm{supp}(W) = \mathbb{R}) \rightarrow (\forall x_{0}, \mathcal{D}'(x_{0}) = \mathcal{D})$.

Note that %For completeness,
%with the model defined in 
%we 
%define one function that implements the entire 
the bit-mask operation at the 
\verb|indexing| %``Indexing'' 
stage (see Fig.~\ref{fig:LUTsystemmodel})  is $M : (\mathcal{Y}, \mathbb{B}^{bN}) \rightarrow \mathcal{D}$ which produces $\mathbf{d} = M(\mathbf{y}, \boldsymbol{q})$
%in accordance with the element-wise definitions given in
following (\ref{eq:decimalexpansiondn}) and (\ref{eq:dyadicexpansionyn}). 
%This function allows us to express
It follows that $\hat{x}(\mathbf{y}) = \hat{x}(M(\mathbf{y},\boldsymbol{q})) = \hat{x}(\mathbf{d}|\boldsymbol{q}).$
%and will also be used later in (\ref{eq:HPImapping}) and Alg.~\ref{alg:montecarlohpi}.

% express I(x0) for d
The Fisher Information for estimating  input $x_{0}$ is % when using bit-mask $\boldsymbol{q}$ is given by:
\begin{align}
    I(x_{0} | \boldsymbol{q}) &= \mathbb{E}_{\mathbf{d} | x_{0}, \boldsymbol{q}}\left[ \left( \frac{\partial \ln ( p(\mathbf{d}|x_{0}, \boldsymbol{q}) )}{\partial x_{0}} \right)^{2} \right]\nonumber\\
    &= \sum_{\mathbf{d} \in \mathcal{D}'(x_{0})} \frac{1}{p(\mathbf{d}|x_{0},\boldsymbol{q})} \cdot \left(\frac{\partial p(\mathbf{d}|x_{0},\boldsymbol{q})}{\partial x_{0}} \right)^{2}
\end{align}
% where $\mathcal{D}$ %$ = \mathcal{I}_{\beta}^{N}$ 
% is the set of all possible bit-masked indexing sequences. $\mathcal{D}$ is strictly a function of the bit-mask $\boldsymbol{q}$, but for notational convenience it is implied that $\mathcal{D} \equiv \mathcal{D}(\boldsymbol{q})$.
where $\frac{\partial p(\mathbf{d}|x_{0},\boldsymbol{q})}{\partial x_{0}}=$
%As expected:
%Since $y_n$ are conditionally independent??, it follows that:
\begin{multline}
    = \idotsint_{\mathbb{R}^{K}} \frac{\partial p(\boldsymbol{\kappa} | x_{0})}{\partial x_{0}} \cdot \left(\prod_{n=-N+1}^{0} p(d_{n} | \boldsymbol{\kappa}, \boldsymbol{q_{n}})\right) d\boldsymbol{\kappa}
\end{multline}

\subsection{Bit Mask Optimization}
% greedy algorithm to determine q*
The determination of optimal bit-mask $\boldsymbol{q}^{*}$  depends on the metric the LUT intends to optimize. 
% !!! here we talk about metrics and our methodology wrt to choosing and evaluating bit-masks: why SFDR is hard to optimize directly and how we adopt an intermediary measure, why we want an analytical heuristic to pre-evaluate instead of a data-driven one, etc
Note  that  $\beta = bN \rightarrow \boldsymbol{q} = \boldsymbol{1} \rightarrow \mathbf{d} = \mathbf{y}$ is equivalent to no bit-masking. 
% two special cases:and $\beta = 0 \rightarrow \boldsymbol{q} = \boldsymbol{0} \rightarrow \mathbf{d} = \mathbf{0}$ is useless, 
Hence,  we study $1 \leq \beta \leq bN-1$.
The optimal bit-mask as:
\begin{align}\label{eq:qstardefn}
    \boldsymbol{q}^{*}(\beta) = \arg \min_{\boldsymbol{q}}&\; H(\boldsymbol{q})\\
    \mathrm{s.t.}&\; \left\| \boldsymbol{q} \right\|_{0} = \beta \nonumber\\
    &\; \boldsymbol{q} \in \mathbb{B}^{bN}\nonumber
\end{align}
where  $H : \mathbb{B}^{bN} \rightarrow \mathbb{R}$ is the  metric.

%\subsubsection{Heuristic Selection Criteria}
%Candidates for bit-mask design heuristics are limited by several constraints. 
% The truly optimal approach for selecting a bit-mask is to forego heuristics altogether 
% %, instead 
% by directly evaluating the desired metric on a pre-determined dataset for each possible bit-mask then choosing the best-performing one by brute-force search. Some metrics actually require this brute-force approach to optimize directly, like SFDR which can only be computed numerically by generating data and computing its frequency spectrum. 
While brute force optimization for  bit-masks when $|\mathbb{B}^{bN}| = 2^{bN}$ is not too large is  possible, it is undesirable for two practical reasons:
\begin{enumerate}
    \item The exponential search is exceedingly   demanding to evaluate for large values of $N,$ a problem which we address by proposing a greedy algorithm in Sec.~\ref{sec:greedybitmasksec}. 
    \item Any method that relies on metric evaluation using collected or simulated data to calibrate the bit-mask is prone to experimental error. Sources of such error include limited sample size (high measurement variance), model mismatch,
% (data includes unexpected interference), 
and outlier events.
% (random chance of low-probability noise sequences).
\end{enumerate} 
%Given the stochastic nature of the modeled inputs these are non-trivial issues, motivating 
An analytical approach to design the bit-mask without reliance on training data is proposed here. 
%This prevents us from optimizing for metrics like 
Nevertheless, such an approach is not tractable for the SFDR metric. % . prompting us to
Hence, instead, we optimize for the MSE and subsequently evaluate its impact on SFDR. Comparison to the brute-force bit-mask search method is included when appropriate as a reference (denoted ``All'' in the legend).

\subsubsection{Bit-Mask Heuristics}
We propose, justify, and evaluate three alternative metrics. First, the data-informed term of the Bayesian Cramer Rao Bound (BCRB) decomposition as described in \cite{crafts_bayesian_2024}, which lower-bounds the MSE of the MMSE estimator:
\begin{align}
    H_{1}(\boldsymbol{q}) &\triangleq -\mathbb{E}_{X_{0}}\! \left[I(X_{0} | \boldsymbol{q}) \right]\\
    &= -\int_{\mathbb{R}} p(x_{0}) \nonumber\\
    &\cdot \sum_{\mathbf{d} \in \mathcal{D}'(x_{0})} \frac{1}{p(\mathbf{d}|x_{0},\boldsymbol{q})} \cdot \left(\frac{\partial p(\mathbf{d}|x_{0},\boldsymbol{q})}{\partial x_{0}} \right)^{2} dx_{0}\nonumber
\end{align}
% \begin{align}
%     %\boldsymbol{q}^{*}(x_{0}) &= \arg \max_{\boldsymbol{q}} I(x_{0}, \boldsymbol{q})\\
%     %\boldsymbol{q}^{*} &= \arg \max_{\boldsymbol{q}} \int_{\mathbb{R}} p(x_{0}) \cdot I(x_{0}, \boldsymbol{q}) dx_{0}
%     \boldsymbol{q}^{*}(\beta) = \arg \max_{\boldsymbol{q}}&\; \mathbb{E}_{x_{0}}\! \left[I(x_{0} | \boldsymbol{q}) \right]\\
%     \mathrm{s.t.}&\; \left\| \boldsymbol{q} \right\|_{0} = \beta \nonumber
% \end{align}
Second, the expectation over $x_{0}$ of the CRLB, where the CRLB bounds the variance of the Minimum-Variance Unbiased Estimator (as per \cite{kay1993fundamentals}):
\begin{align}
    H_{2}(\boldsymbol{q}) &\triangleq \mathbb{E}_{X_{0}}\! \left[I^{-1}(X_{0} | \boldsymbol{q}) \right]\\
    &= \int_{\mathbb{R}} p(x_{0}) \nonumber\\
    &\cdot \left(\sum_{\mathbf{d} \in \mathcal{D}'(x_{0})} \frac{1}{p(\mathbf{d}|x_{0},\boldsymbol{q})} \cdot \left(\frac{\partial p(\mathbf{d}|x_{0},\boldsymbol{q})}{\partial x_{0}} \right)^{2}\right)^{-1} dx_{0}\nonumber
\end{align}
%  \nonumber\\ &\hspace{0.5cm}
% \begin{align}
%     \boldsymbol{q}^{*}(\beta) = \arg \min_{\boldsymbol{q}}&\;  \mathbb{E}_{x_{0}}\! \left[I^{-1}(x_{0} | \boldsymbol{q}) \right]\\
%     \mathrm{s.t.}&\; \left\| \boldsymbol{q} \right\|_{0} = \beta \nonumber
% \end{align}
Third, the expected MSE of the MMSE estimator %of the high-resolution estimate $\hat{x}_{0}$% when using the MMSE estimator (derived in Appendix~\ref{sec:H3derivation}):
\begin{align}
    H_{3}(\boldsymbol{q}) &\triangleq 
    %\left\|\hat{\mathbf{x}}_{\mathrm{MMSE}}(\boldsymbol{q}) - \mathbf{x}\right\|_{2}^{2}\\
    \mathbb{E}_{\hat{X}_{0},X_{0}}\left[\left(\hat{X}_{0} - X_{0}\right)^{2} \Big| \boldsymbol{q}\right] \\
    &= \int_{\mathbb{R}} p(x_{0}) \cdot \sum_{\mathbf{d} \in \mathcal{D}'(x_{0})} p(\mathbf{d}|x_{0},\boldsymbol{q})\nonumber\\
    &\hspace{0.5cm} \cdot \left(\frac{\int_{-\infty}^{\infty}x_{0} \cdot p(x_{0}) \cdot p(\mathbf{d} | x_{0}, \boldsymbol{q}) dx_{0}}{\int_{-\infty}^{\infty} p(x_{0}) \cdot p(\mathbf{d} | x_{0}, \boldsymbol{q}) dx_{0}} - x_{0}\right)^{2} dx_{0}\nonumber
\end{align}

Last, as a control group we study the naive sequential indexing of the bit-mask (denoted $H_{0}$), defined directly as: 
\begin{equation}
    %\arg \min_{\boldsymbol{q}} H_{0} = 
    q_{[i]}(\beta) = \left\{\begin{matrix}
    0, & i < bN-\beta+1 \\
    1, & i \geq bN-\beta+1 \\
    \end{matrix}\right. 
\end{equation}

%The set of possible bit-masks $\boldsymbol{q} \in \mathbb{B}^{bN}$ with $\left\|\boldsymbol{q}\right\| = \beta$ is of cardinality $?$. 

\subsection{Greedy Computation}\label{sec:greedybitmasksec}
% Since we cannot, in general, guarantee convexity of 
% any of the proposed heuristic metrics
% %$I(x_{0} | \boldsymbol{q})$ 
% % we would need to evaluate each possible bit-mask to determine the optimal one, requiring
% brute force search woud require $\binom{bN}{\beta} = \frac{(bN)!}{\beta ! (bN - \beta)!}$ computations with combinatorial time complexity over $\beta$. 
We propose an iterative Greedy Algorithm for selecting bits in the mask 
%, described in Algorithm~\ref{alg:greedybm} 
% which minimizes an arbitrary heuristic function $H,$ 
as follows: 
%completing $\beta$ iterations, 
initialize $\boldsymbol{q}^{(0)} = \boldsymbol{0}, \mathcal{Q}^{(0)} = \mathbb{B}^{bN}, \mathcal{J}^{(0)} = \emptyset$, and at, each iteration, $t = \{1, \cdots, \beta\}$ apply the sequential update equations: 
\begin{align}
    \mathcal{Q}^{(t)} &= \{\boldsymbol{q} \in \mathbb{B}^{bN}\; |\; \forall j \in \mathcal{J}^{(t-1)}, q_{[j]} = 1\; \wedge\; \left\| \boldsymbol{q} \right\|_{0} = t\}\nonumber\\
    \boldsymbol{q}^{(t)} &= \arg \min_{\boldsymbol{q} \in \mathcal{Q}^{(t)}} H(\boldsymbol{q})\\
    \mathcal{J}^{(t)} &= \{j\; |\; q_{[j]}^{(t)} = 1\}\nonumber
\end{align}
% In this way, 
The algorithm only changes one bit per iteration and only has to test any remaining 0 entries in the next iteration, greatly reducing the number of bit-masks to be evaluated. It also procedurally generates the greedy bit-masks for all smaller values of $\beta$ in the process as $\boldsymbol{q}^{(t)}$. This procedure is formalized in Algorithm~\ref{alg:greedybm}.
\begin{algorithm}[h]
    \caption{Greedy Bit-Mask Selection Algorithm}\label{alg:greedybm}
    \begin{algorithmic}
        \Require $\beta \in \{ 0, \cdots, bN\}, H : \mathbb{B}^{bN} \rightarrow \mathbb{R}$
        \Ensure $\boldsymbol{q}^{(\beta)} \in \mathbb{B}^{bN}, \left\| \boldsymbol{q}^{(\beta)} \right\|_{0} = \beta$
        \State $\boldsymbol{q}^{(0)} \gets \boldsymbol{0}$
        %\State{$\boldsymbol{s} \gets \boldsymbol{q}$}
        %\State $\mathcal{Q}^{(0)} \gets \mathbb{B}^{bN}$
        \State $\mathcal{J}^{(0)} \gets \emptyset$
        \State $h^{*} \gets \infty$
        \For{$t = 1, \cdots, \beta$}
            \For{$j \in \{1, \cdots, bN\} \setminus \mathcal{J}^{(t-1)}$}
                \State $\boldsymbol{q} \gets \boldsymbol{q}^{(t-1)}$
                \State $q_{[j]} \gets 1$
                \If{$H(\boldsymbol{q}) \leq h^{*}$}
                    \State $\boldsymbol{q}^{(t)} \gets \boldsymbol{q}$
                    \State $j^{*} \gets j$
                    \State $h^{*} \gets H(\boldsymbol{q})$
                \EndIf
            \EndFor
            %\State{$\boldsymbol{q} \gets \boldsymbol{s}$}
            \State{$\mathcal{J}^{(t)} \gets \mathcal{J}^{(t-1)} \cup \{j^{*}\}$}
        \EndFor
        \State \Return $\boldsymbol{q}^{(\beta)}$
    \end{algorithmic}
\end{algorithm}
% \begin{algorithm}[h]
%     \caption{Greedy Bit-Mask Selection Algorithm}\label{alg:greedybm}
%     \begin{algorithmic}
%         \Require $\beta \in \{ 0, \cdots, bN\}, H : \mathbb{B}^{bN} \rightarrow \mathbb{R}$
%         \Ensure $\boldsymbol{q} \in \mathbb{B}^{bN}, \left\| \boldsymbol{q} \right\|_{0} = \beta$
%         \State $\boldsymbol{q} \gets \boldsymbol{0}$
%         \State{$\boldsymbol{s} \gets \boldsymbol{q}$}
%         \State $\mathcal{I} \gets \emptyset$
%         \State $h^{*} \gets \infty$
%         \For{$i = 1, \cdots, \beta$}
%             %\State{$\boldsymbol{s} \gets \boldsymbol{q}$}
%             \For{$j \in \{1, \cdots, bN\} \setminus \mathcal{I}$}
%                 \State $\boldsymbol{t} \gets \boldsymbol{q}$
%                 \State $t_{[j]} \gets 1$
%                 \If{$H(\boldsymbol{t}) \leq h^{*}$}
%                     \State $\boldsymbol{s} \gets \boldsymbol{t}$
%                     \State $\mathcal{S} \gets \{j\}$
%                     \State $h^{*} \gets H(\boldsymbol{t})$
%                 \EndIf
%             \EndFor
%             \State{$\boldsymbol{q} \gets \boldsymbol{s}$}
%             \State{$\mathcal{I} \gets \mathcal{I} \cup \mathcal{S}$}
%         \EndFor
%         \State \Return $\boldsymbol{q}$
%     \end{algorithmic}
% \end{algorithm}
Note that due to the $\leq$ operator, implied tie-break criteria favors bits closer to $n=0$ and %within each $n$ 
  favors bits closer to the LSB.

This greedy algorithm requires only $\sum_{i=0}^{\beta - 1} (bN - i) = - \frac{1}{2}\beta^{2} + \left(bN + \frac{1}{2}\right)\beta$ bit-mask evaluations and, hence, is of  polynomial time complexity.
% $\mathcal{O}(\beta^{2})$.
%In both the greedy and optimal cases, each tested bit-mask requires $2^{\beta}$ computations of $p(\mathbf{d} | x_{0}, \boldsymbol{q})$.  % cut because beta is the final value in greedy, not the instantaneous |qv|_{0} value...
% The performance impact of using the greedy heuristic will also be evaluated.

% \begin{center}
%     \begin{tabular}{|c|c|}
%         \hline
%         Optimal & Greedy \\
%         \hline
%         $(bN)! / (\beta!(bN-\beta)!)$ & $- \frac{1}{2}\beta^{2} + \left(bN + \frac{1}{2}\right)\beta$\\
%         \hline
%     \end{tabular}
% \end{center}

%X fisher information for tone with bit-mask

% ...results...
\subsection{Bit-Mask Selection: Numerical Results}
% include parameters from previous paper
%Simulation setup parameters are the same as those described in Sec.~\ref{sec:simsetup}. 
All simulated results are computed by evaluating $\hat{x}_{\mathrm{MMSE}}(\mathbf{d}|\boldsymbol{q})$ as per (\ref{eq:mmseformulationdv})  and in Sec.~\ref{sec:simsetup} (without  dithering or requantization unless otherwise stated explicitly).

For ease of notation, estimation with bit-masks solved using the optimal combinatorial method in (\ref{eq:qstardefn}) for heuristic $H_{i}$ are referred to as $H_{i}^{*}$ bit-masks while those using the sub-optimal greedy method in Alg.~\ref{alg:greedybm} are  $H_{i}^{G}$ bit-masks. 

\begin{figure}[h]
    \centering
    \includegraphics[width=0.48\textwidth]{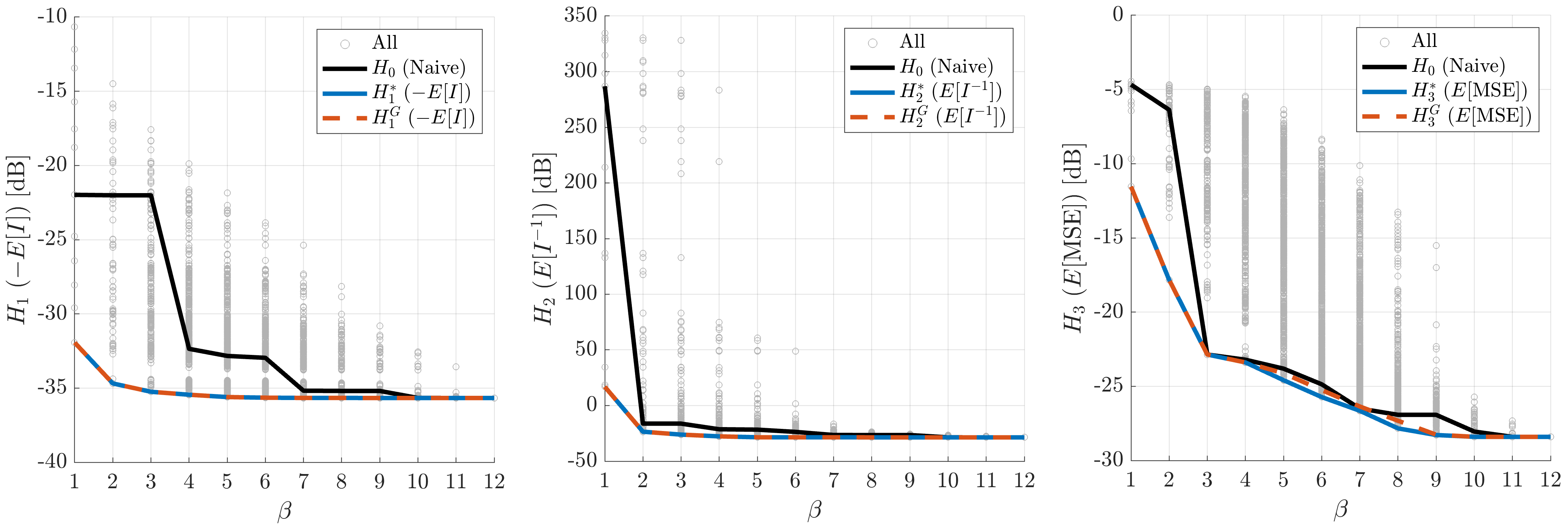}
    \caption{Heuristic Performance of Bit-Masks Trained for Simulated Quantized Tone with $N=4$}
    \label{fig:HPerfBeta}
\end{figure}
Training results in Fig.~\ref{fig:HPerfBeta} illustrate several insights. First, computation of $H_{2}$ is numerically unstable due to the inverse in its definition resulting in outliers with extremely high evaluations making it unreliable. Second, the optimal bit-mask achieves almost identical evaluations to the greedy bit-masks for $H_{1}$ and $H_{2}$, but differs by a non-negligible margin for $H_{3}$. Third, in all three heuristics and for all tested $\beta$ values there is an improvement over using the naive sequential bit-mask. 
Nevertheless, %when comparing the superset of all possible bit-masks each independently evaluated on these metrics, 
the naive method is typically not far off from the optimal bit-mask, and is usually at least better than average.
%(as $\lim_{x_{0} \rightarrow \pm A} p(x_{0}) = \infty$)

\begin{figure}[h]
    \centering
    \includegraphics[width=0.48\textwidth]{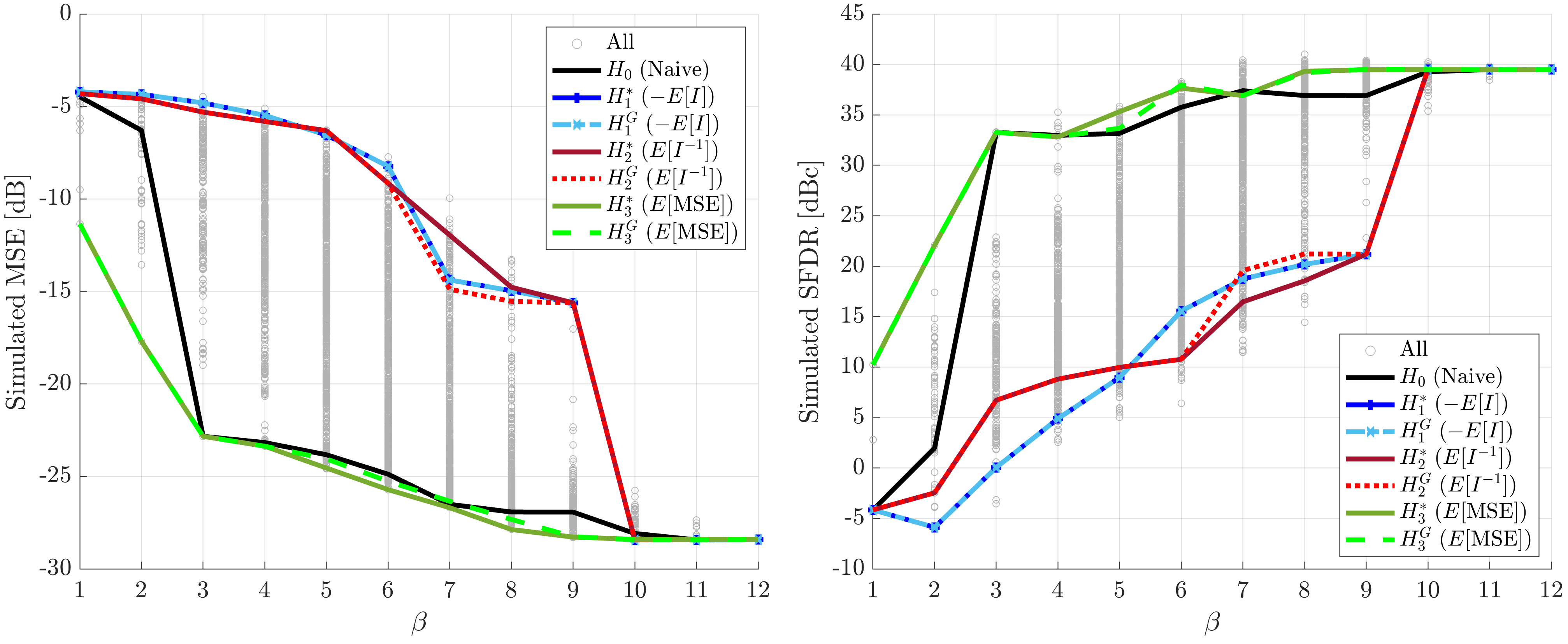}
    \caption{Evaluation of $\hat{x}_{\mathrm{MMSE}}$ for Simulated Quantized Tone After Bit-Masking and Estimation with $N=4, \beta = \{1, \cdots, bN\}$}
    \label{fig:HEvalBeta}
\end{figure}
Next we test the performance of the optimized bit-masks on two  key metrics: MSE and SFDR. The results shown in Fig.~\ref{fig:HEvalBeta} reveal further insights. First: the heuristics $H_{1}$ and $H_{2}$ produce some of the worst-performing bit-masks, with $H_{3}$ and $H_{0}$ being the only consistent high-performers. One likely reason for this is that Cramer-Rao-style bounds are only tight for  Gaussian posterior distributions~\cite{trees_detection_2013}, which our input signal does not satisfy. 
%This strongly supports the use of $H_{3}$ for bit-mask determination over all other heuristics.
%especially since $H_{3}$ is a direct evaluation rather than a (potentially loose) bound. 
%Among those trained on the $H_{3}$ heuristic, 
Further, the optimal and greedy bit-masks perform similarly with a narrow margin of difference 
%(within $1$ dB MSE) 
while both consistently out-perform the naive $H_{0}$ bit-mask ($\approx 1$ dB MSE improvement). 
%Despite only being trained for MSE,
Further,  $H_{3}$ bit-masks (both greedy and optimal) also perform well on SFDR and are both typically within a few dBc of the optimal value.
In the rest of the paper we use  $H_{3}$ heuristic.
%as determined by the superset of all possible bit-masks. 
%When compared to the superset of all possible bit-masks the $H_{3}$-optimal masks are also optimal in the mean-square sense when simulated, and close to optimal for SFDR (within a few dBc).

\begin{figure}[h]
    \centering
    \includegraphics[width=0.48\textwidth]{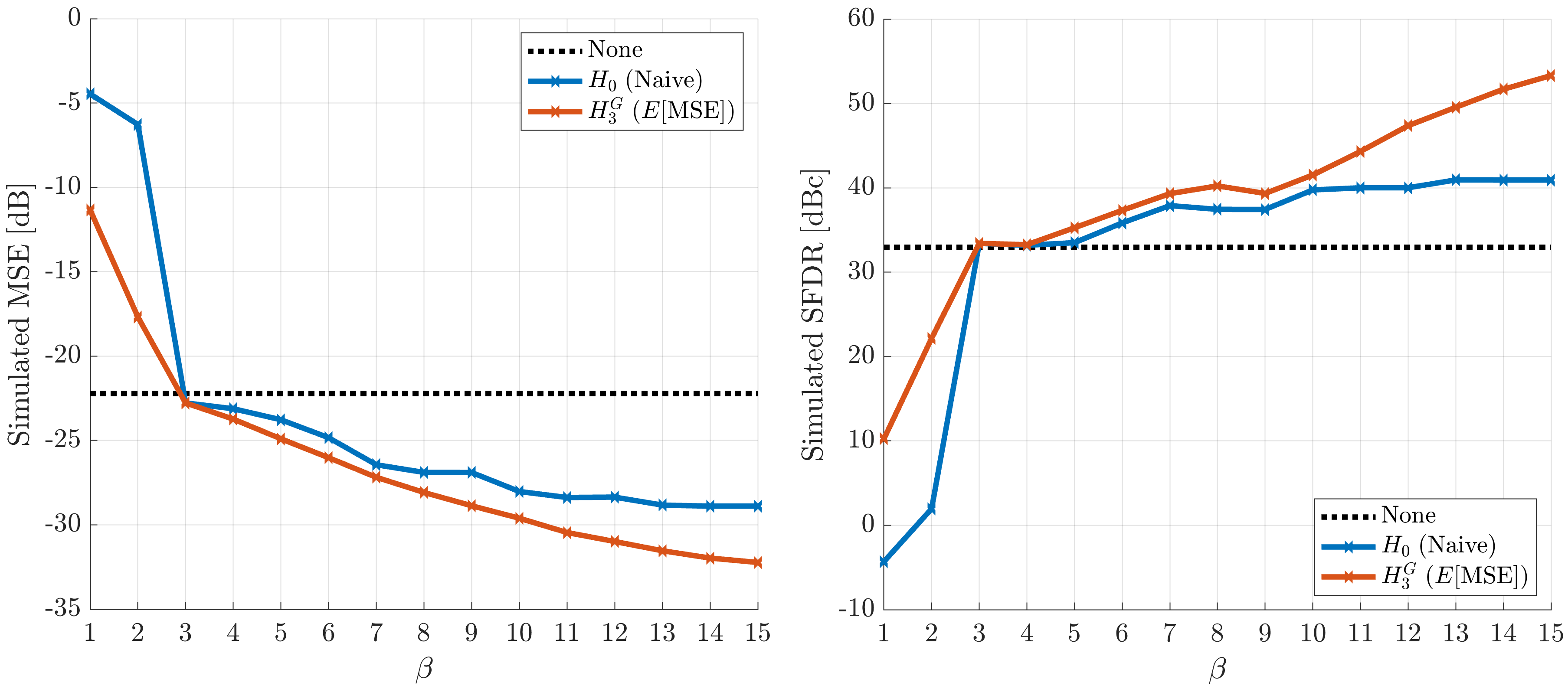}
    \caption{Evaluation of $\hat{x}_{\mathrm{MMSE}}$ for Simulated Quantized Tone Estimated Using Greedy $H_{3}$ and Naive $H_{0}$ Bit-Masks with $N=10, \beta = \{1, \cdots, bN/2\}$. ``None'' Evaluates Input $Q(x+w)$ Directly (No Estimator)}
    \label{fig:GH3EvalBeta}
\end{figure}
%Armed with this new knowledge, 
The relative performance of the greedy $H_{3}$ heuristic and the naive $H_{0}$ sequential bit-masks is evaluated with greater rigor by training on a much longer $N=10$ window and constraining $\beta \leq bN/2$ to compare only within the high degree-of-freedom training region. 
Results  in Fig.~\ref{fig:GH3EvalBeta} illustrate that,
%despite the previously-discovered performance penalty associated with the greedy optimization method, 
the trained $H_{3}^{G}$ bit-masks still consistently outperform the $H_{0}$ naive bit-masks in both MSE and SFDR for all values of $\beta$. In this example the index size $\beta = bN/2 = 15$  produces  gains of $>3$ dB MSE and of $>12$ dBc SFDR.
\begin{figure}[h]
    \centering
    \includegraphics[width=0.48\textwidth]{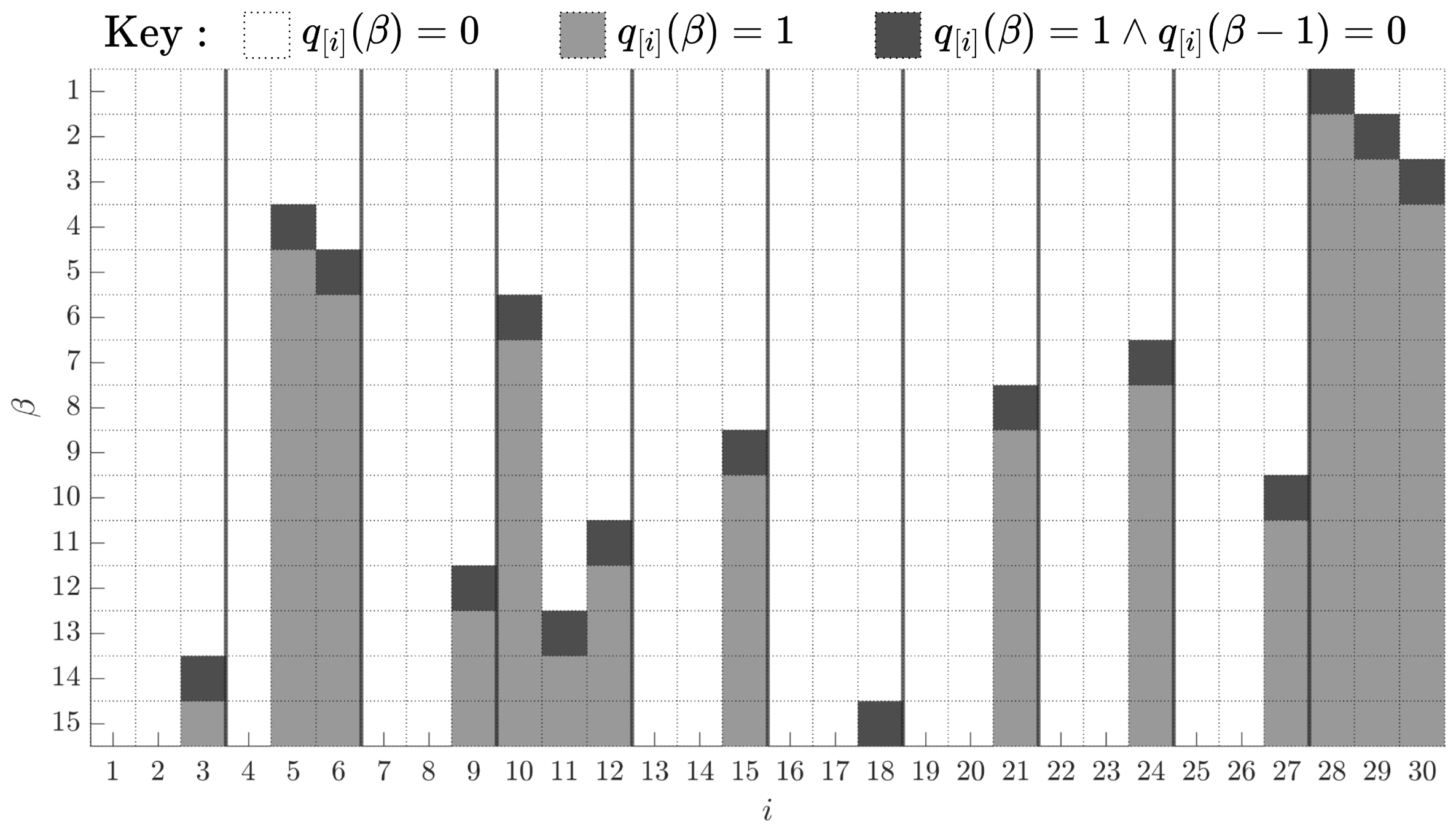}
    \caption{Bit-Mask Optimized Using Greedy $H_{3}$ Heuristic for $N=10$ (Solid Vertical Lines Separate $n = \{-9, \cdots, 0\}$ with Big-Endian Bits Within Each $\boldsymbol{q_{n}}$ [MSB $\rightarrow$ LSB])}
    \label{fig:GH3qvSelection}
\end{figure}
The bits chosen for one such bit-mask are illustrated graphically in Fig.~\ref{fig:GH3qvSelection}. Notably this bit-mask differs significantly from the naive one, as aggregating LSBs across different samples is typically favored over choosing multiple bits in the same sample.

\begin{figure}[h]
    \centering
    \includegraphics[width=0.48\textwidth]{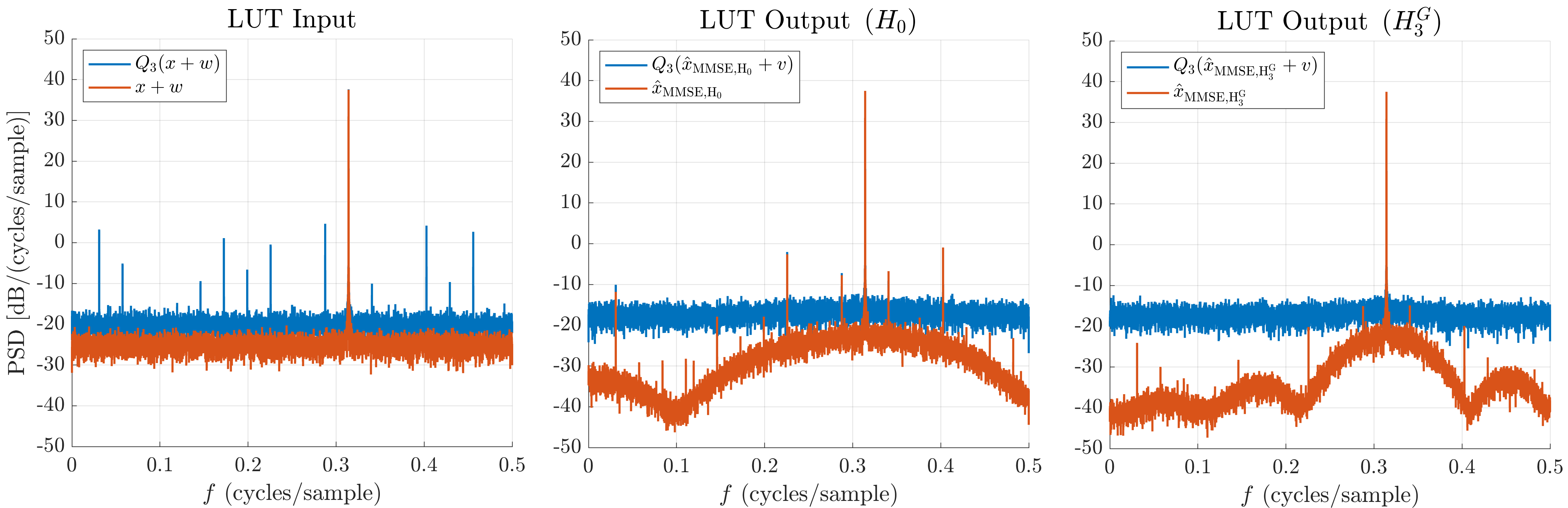}
    \caption{PSD Comparison for Quantized Tone After High-Resolution Estimation ($N = 10, \beta = 15$) and After Low-Resolution Requantization with Post-Table Dithering ($\alpha=1$). Results Shown for Input Signal (Left), Naive Sequential $H_{0}$ Indexing (Center), and Greedy $H_{3}^{G}$ Indexing (Right)}
    \label{fig:qvPSDcomparison}
\end{figure}
%The numerical results in
%Fig.~\ref{fig:GH3EvalBeta} illustrates a PSD comparison for one example bit-mask in
Fig.~\ref{fig:qvPSDcomparison} illustrates that 
% The results show that despite a fixed total index size $\beta=15$, the use of
the $H_{3}^{G}$-optimized bit-mask can dramatically improve SFDR %at the output of the \verb|estimation| stage 
compared to the naive $H_{0}$ sequential indexing. 
This is achieved by both reducing quantization spurs near the frequency of interest and attenuating quantization noise throughout the power spectrum. 
%As a result of this improved high-resolution estimate, 
The overall LUT SFDR gain is  preserved %after post-table since
%the noise floor is increased without producing
 since no new harmonic spurs are produced after  dithering and requantization. Because memory size is a function of $\beta$, both $H_{0}$ and $H_{3}^{G}$ LUTs tested require the same total memory making this an apt comparison for memory-constrained systems.% showing an SFDR performance gain by appropriate bit-masking.

%The discrepancy between the ``visual'' SFDR obtained by inspecting Fig.~\ref{fig:qvPSDcomparison} being several dB better than the calculated SFDR results presented in Fig.~\ref{fig:GH3EvalBeta} is explained by the strict SFDR definition in (\ref{eq:sfdrdefn}) and commentary in the paragraph that follows it.

\section{Memory Optimization}\label{sec:memoryopt}
Recall from (\ref{eq:memorysizedefn}) that the memory size of a LUT in bits is defined as $\rho \cdot L$, where $\rho > b$ for a post-table dithering architecture as per Table~\ref{tbl:ditheringcomparison}. %In practice LUTs must adhere to a total memory size constraint imparted by the implementation hardware. 
Reducing the memory size requires either lowering $\rho$ directly or decreasing $L$. When naively training the LUT for all possible indexing sequences $L = |\mathcal{Y}| = 2^{bN}$. Bit-masking reduces this value to $L = |\mathcal{D}| = 2^{\beta} < 2^{bN}$. We propose an additional method to further reduce $L$: high-probability indexing.

\subsection{High-Probability Indexing}\label{sec:hpisec}
For a given bit-mask $\boldsymbol{q}$, define the associated High-Probability Indexing (HPI) set $\mathcal{D}_{\epsilon}(\boldsymbol{q})$ as:
\begin{align}\label{eq:HPIdefn}
    \mathcal{D}_{\epsilon}(\boldsymbol{q}) \triangleq \arg \min_{\mathcal{D}}&\; |\mathcal{D}| \\
    \mathrm{s.t.}&\; \sum_{\mathbf{d} \in \mathcal{D}} p(\mathbf{d}|\boldsymbol{q}) \geq \epsilon \nonumber\\
    &\; \mathcal{D} \subseteq \mathcal{I}_{b}^{N} \nonumber
    %\mathcal{I}_{\left\| \boldsymbol{q} \right\|_{0}}^{N} \nonumber
\end{align}
%where $\beta = \left\| \boldsymbol{q} \right\|_{0}$. 
where
\begin{equation}\label{eq:pdvcondq}
    p(\mathbf{d}|\boldsymbol{q}) = \int_{\mathbb{R}} p(x_{0}) \cdot p(\mathbf{d}|x_{0},\boldsymbol{q}) \;dx_{0}
\end{equation}
and $p(\mathbf{d}|x_{0},\boldsymbol{q})$ is given by (\ref{eq:pdcondx0q}).
%Based on this definition we can see that 
%While $\mathcal{D}$ is strictly a function of $\boldsymbol{q}$, 
The expression for $\mathcal{D}_{\epsilon}(\boldsymbol{q})$ does not depend on $x_{0}$ directly but it does depend on $p(x_{0})$ and $p(\mathbf{d}|x_{0},\boldsymbol{q})$, which is itself a function of $p(\boldsymbol{\kappa}|x_{0})$ and $p(y_{n} | \boldsymbol{\kappa})$. Consequently the HPI set depends on the parametric model for the input signal $x_{n}$, the prior distributions used for its parameters $p(\boldsymbol{\kappa})$, and the input quantizer's transfer function. These additional variables are omitted from the notation for clarity but are necessary for computation.

\begin{figure}[h]
    \centering
    \includegraphics[width=0.48\textwidth]{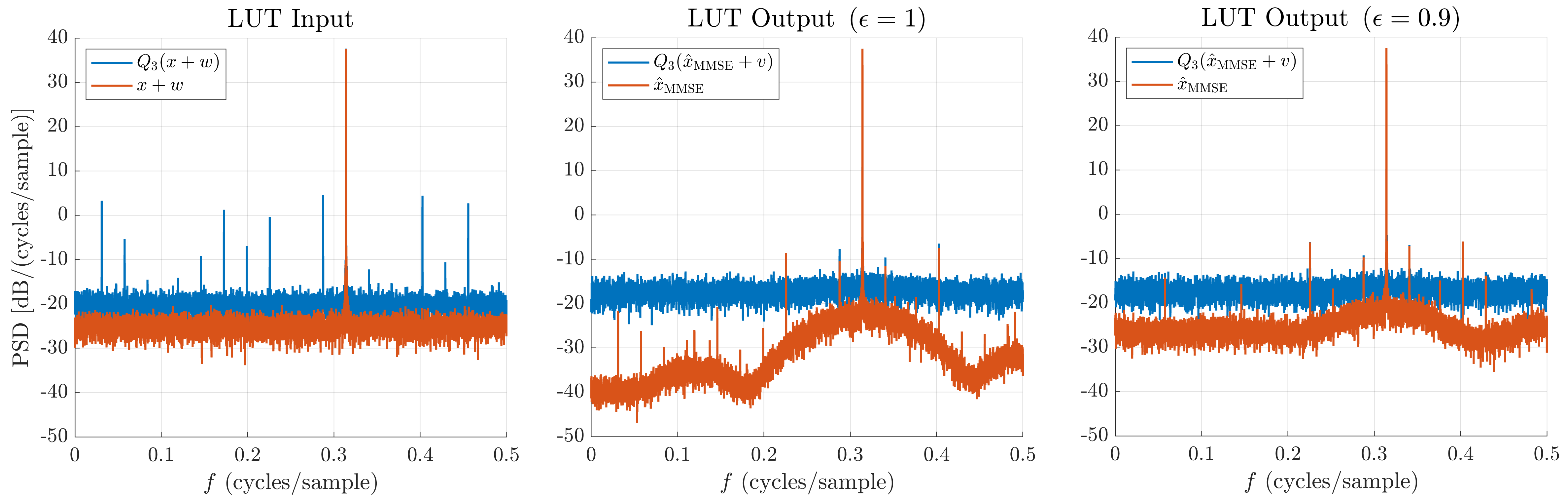}
    \caption{PSD Comparison for Quantized Tone (Left) Estimated and Dithered+Requantized Using Full Indexing $\epsilon=1$ (Center) and High-Probability Indexing $\epsilon=0.9$ (Right) with $N = 7$}
    \label{fig:HPIPSDComparison}
\end{figure}

$\mathcal{D}_{\epsilon}$ is the smallest set containing at least $\epsilon$ proportion of the total indexing probability. Hence, a LUT trained for only this set of indices is expected to be able to correct $\epsilon$ proportion of inputs. 
% We assume that for any input indexing sequence not present in the set, the LUT will be bypassed and output the current digital input value $y_{0}$.
For any index not in the set , the LUT  output is assigned to the current digital input value $y_{0}$.
% To fit this concept into the model for the \verb|indexing| stage in Fig.~\ref{fig:LUTsystemmodel}, 
Hence, we express the HPI LUT mapping as:
\begin{equation}\label{eq:HPImapping}
    \ell_{\epsilon}(\mathbf{y}|\boldsymbol{q}) = \left\{\begin{matrix}
\ell(\mathbf{y}), & M(\mathbf{y},\boldsymbol{q}) \in \mathcal{D}_{\epsilon}(\boldsymbol{q}) \\
y_{0}, & M(\mathbf{y},\boldsymbol{q}) \notin \mathcal{D}_{\epsilon}(\boldsymbol{q}) \\
\end{matrix}\right. 
\end{equation}
in terms of an arbitrary non-HPI LUT mapping $\ell(\mathbf{y})$. 
%where $d : (\mathcal{Y}, \mathbb{B}^{bN}) \rightarrow \mathcal{D}$ is the function implementing the bit-mask operation on each sample as per (\ref{eq:decimalexpansiondn}).
%\subsubsection{Simulated Results}
% For the sake of general applicability 
We also illustrate the results for $\mathbf{y} \in \mathcal{Y} = \mathcal{I}_{b}^{N}$ since it represents the special case $\mathcal{Y}_\epsilon = \mathcal{D}_\epsilon(\boldsymbol{1})$ and any results over $\mathcal{D}$ will be highly dependent on $\boldsymbol{q}$ and therefore on $\beta$. Note that  $|\mathcal{Y}_{1}| = |\mathcal{Y}| = 2^{bN}$.
%$\mathcal{Y}_{\epsilon}$ is defined in the same way as $\mathcal{D}_{\epsilon}(\boldsymbol{1})$, as per (\ref{eq:HPIdefn}).
% information here about distribution of pyv
\begin{figure}[h]
    \centering
    \includegraphics[width=0.48\textwidth]{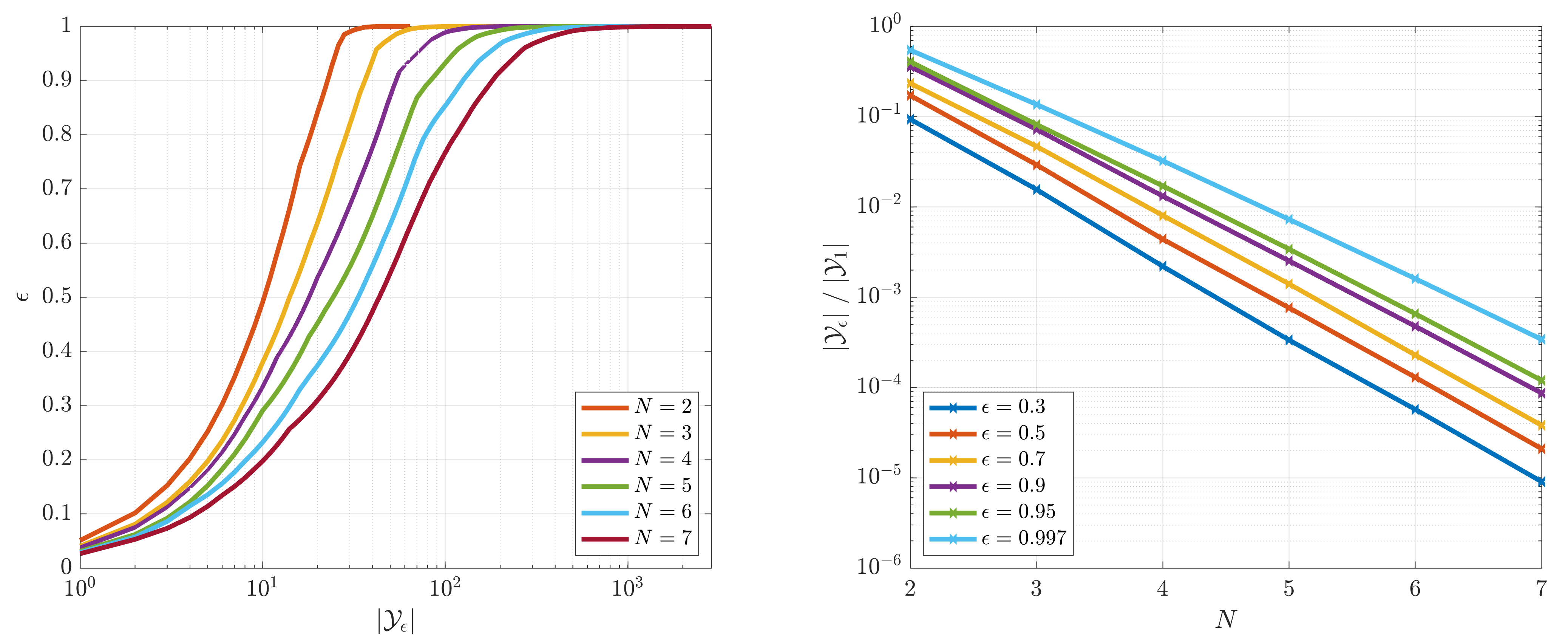}
    \caption{Relative Size of High-Probability Indexing Set for Quantized Tone Over $\epsilon$ and $N$}
    \label{fig:CalYEpsSize}
\end{figure}
As shown by Fig.~\ref{fig:CalYEpsSize}, even a very small total number of elements can produce an $\epsilon$ value very close to 1, indicating that the probability mass in the distribution $p(\mathbf{y})$ is highly concentrated in a small subset of indexing sequences. This is further evidenced by the extremely efficient ratio of the size of the HPI set to the full indexing set, with increasing efficiency over $N$ even for $\epsilon$ very close to 1.
For $N=7$ and $\epsilon=0.9$ the efficiency increases by over four orders of magnitude.  
\begin{figure}[h]
    \centering
    \includegraphics[width=0.4\textwidth]{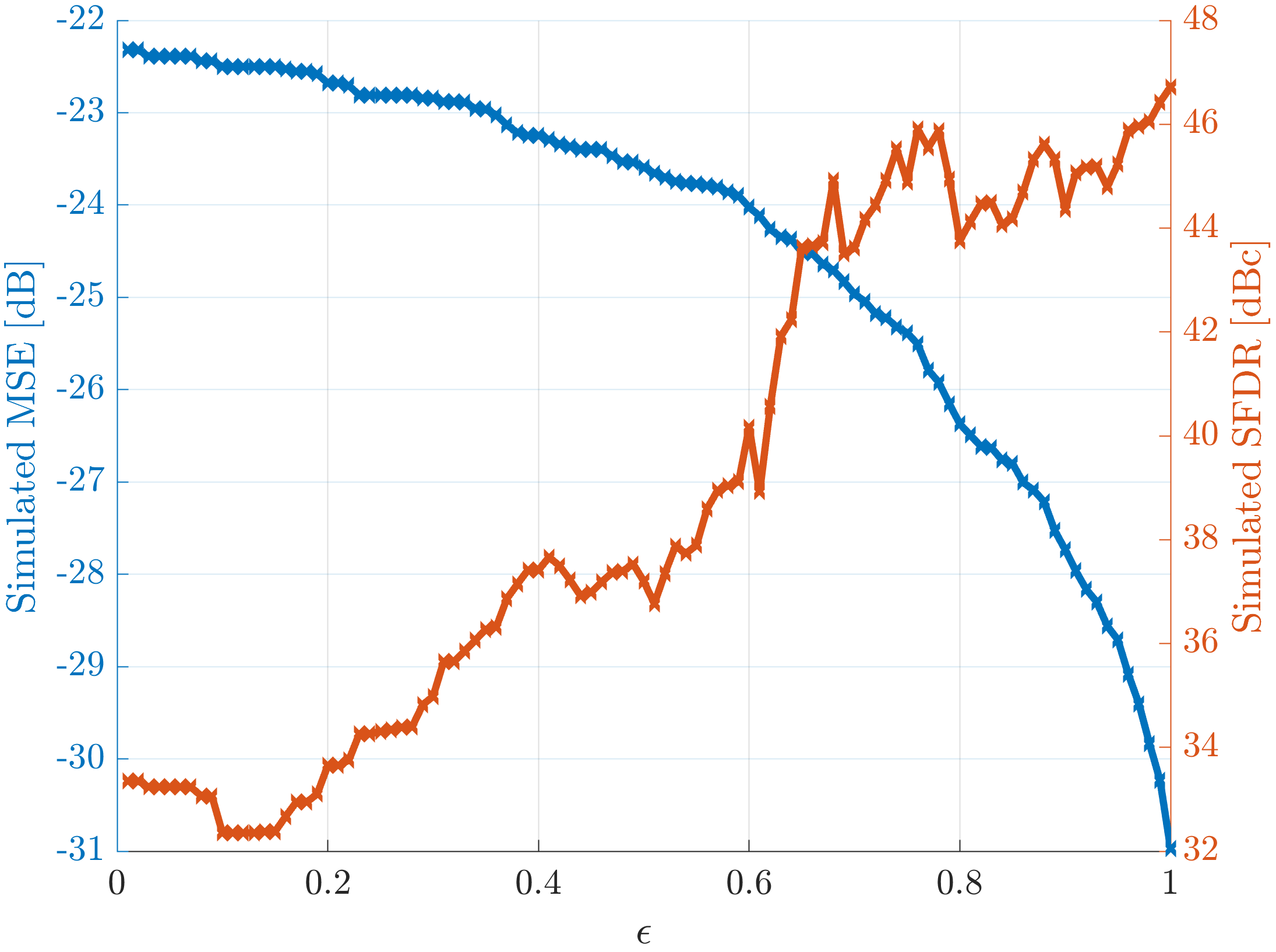}
    \caption{Relative Performance Over $\epsilon$ of Estimate $\hat{x}_{\mathrm{MMSE}}$ After High-Probability Indexing for Simulated Quantized Tone with $N=7$}%, Without Bit-Masking}
    \label{fig:MetricEvaluationOverEps}
\end{figure}
% Despite the large memory gains associated with using $\epsilon < 1$, 
Furthermore,
Fig.~\ref{fig:MetricEvaluationOverEps} evidences that even values as low as $\epsilon = 0.68$ are within 2 dBc SFDR of the maximum value at $\epsilon = 1$. This  result is  illustrated by the PSD  comparison in Fig.~\ref{fig:HPIPSDComparison}, which shows that after the high-resolution \verb|estimation| stage the output spectra ,when using $\epsilon = \{0.9, 1\},$ share almost identical peak spurs while the MSE increase is caused almost exclusively by the increased out-of-band noise. 
% This figure also shows how, even after the \verb|requantization| stage from Fig.~\ref{fig:LUTsystemmodel}, 
Further, we also observe that the SFDR gains exhibit negligible loss
%are maintained almost exactly as no 
since no new spurs appear 
%despite the signal being reduced to 
after requantization to $b=3$-bit resolution.

\subsection{Efficient Approximation of the HPI Set}
Determination of the high-probability indexing set as defined in (\ref{eq:HPIdefn}) requires direct computation of $p(\mathbf{d} | \boldsymbol{q})$ for $|\mathcal{D}| = 2^{\beta}$ values of $\mathbf{d}$ and, thus, exponential time complexity. To mitigate this computational burden, we propose to approximate the HPI set through Monte-Carlo generation of $\Upsilon$ sequences as formalized in Algorithm~\ref{alg:montecarlohpi}.
%we propose to generate $\Upsilon$ sequences of the input $\mathbf{x}$ (each of length $N$) according to its parametric model and use the set of unique indexing sequences encountered by this simulated input as our high-probability indexing set. This process is formalized in Algorithm~\ref{alg:montecarlohpi}.
% \begin{algorithm}[h]
%     \caption{Monte-Carlo HPI Set Approximation}\label{alg:montecarlohpi}
%     \begin{algorithmic}
%         \Require $\Upsilon \in \mathbb{N}, p(\boldsymbol{\kappa}) : \mathbb{R}^{K} \rightarrow [0, 1], p(w) : \mathbb{R} \rightarrow [0,1]$\\
%         $\hspace{12mm}\boldsymbol{q} \in \mathbb{B}^{bN}, \mathbf{x} : \mathbb{R}^{K} \rightarrow \mathbb{R}^{N}, \Call{mask}{} :(\mathcal{Y}, \mathbb{B}^{bN}) \rightarrow \mathcal{D}$
%         \Ensure $|\mathcal{U}| > 0$
%         \State $\mathcal{U} \gets \emptyset$
%         \For{$i = 1, \cdots, \Upsilon$}
%             \State $\boldsymbol{\kappa}_{[i]} \sim p(\boldsymbol{\kappa})$
%             \State $w_{[i]} \sim p(w)$
%             \State $\mathbf{y}_{[i]} \gets Q_{b}(\mathbf{x}(\boldsymbol{\kappa_{[i]}}) + w_{[i]})$
%             \State $\mathbf{d}_{[i]} \gets \Call{mask}{\mathbf{y}_{[i]}, \boldsymbol{q}}$
%             \State{$\mathcal{U} \gets \mathcal{U} \cup \{\mathbf{d}_{[i]}\}$}
%         \EndFor
%         \State \Return $\mathcal{U}$
%     \end{algorithmic}
% \end{algorithm}
\begin{algorithm}[h]
    \caption{Monte-Carlo HPI Set Approximation}\label{alg:montecarlohpi}
    \begin{algorithmic}
        \Require $\Upsilon \in \mathbb{N}, p(\boldsymbol{\kappa}) : \mathbb{R}^{K} \rightarrow [0, 1], p(w) : \mathbb{R} \rightarrow [0,1]$\\
        $\hspace{12mm}\boldsymbol{q} \in \mathbb{B}^{bN}, \mathbf{x} : \mathbb{R}^{K} \rightarrow \mathbb{R}^{N}, M :(\mathcal{Y}, \mathbb{B}^{bN}) \rightarrow \mathcal{D}$
        \Ensure $|\mathcal{U}| > 0$
        \State $\mathcal{U} \gets \emptyset$
        \For{$i = 1, \cdots, \Upsilon$}
            \State $\boldsymbol{\kappa}_{[i]} \sim p(\boldsymbol{\kappa})$
            \State $w_{[i]} \sim p(w)$
            \State $\mathbf{y}_{[i]} \gets Q_{b}(\mathbf{x}(\boldsymbol{\kappa_{[i]}}) + w_{[i]})$
            \State $\mathbf{d}_{[i]} \gets M(\mathbf{y}_{[i]}, \boldsymbol{q})$
            \State{$\mathcal{U} \gets \mathcal{U} \cup \{\mathbf{d}_{[i]}\}$}
        \EndFor
        \State \Return $\mathcal{U}$
    \end{algorithmic}
\end{algorithm}
Generation of the input signal is typically much faster than evaluating $p(\mathbf{d} | \boldsymbol{q})$ according to (\ref{eq:pdvcondq}) directly. %In order for this to be an effective 
In order to utilize this method we must predict how many samples $\Upsilon$ will produce an expected total indexing probability $\epsilon$. That relationship is explored in the following analysis, 
%In practice, generating samples sequentially will result in temporally-correlated realizations of indexing sequences. For analysis we assume independence of each indexing sequence realization, which is approximately true for well-behaved inputs and for large $\Upsilon$.
where $\boldsymbol{q}$ will be omitted for clarity. %from expressions involving $p(\mathbf{d}|\boldsymbol{q})$ as its presence is implied by $p(\mathbf{d})$.
%For notational convenience, it is implied that $p(\mathbf{d}) = p(\mathbf{d} | \boldsymbol{q})$ since $\mathbf{d}$ is only defined for a given $\boldsymbol{q}$.
%We identify each possible bit-masked input sequence by assigning to it an arbitrary unique index: 
We index the set $\mathcal{D}$ with unique subscripts $\mathbf{d}_{[j]} \in \mathcal{D}, j = \{1, \cdots, 2^{\beta}\}$. 
%Now, consider generating $\Upsilon$ indexing sequences by randomly generating $N$-length windows of $\mathbf{x}$ (as per Alg.~\ref{alg:montecarlohpi}). 
Denote the set of all unique indexing sequences $\mathbf{d}$ generated by the Monte-Carlo algorithm as $\mathcal{U} \subseteq \mathcal{D}$. Naturally, both the elements of $\mathcal{U}$ and its size $|\mathcal{U}|$ will be stochastic.

\begin{figure}[h]
    \centering
    \includegraphics[width=0.4\textwidth]{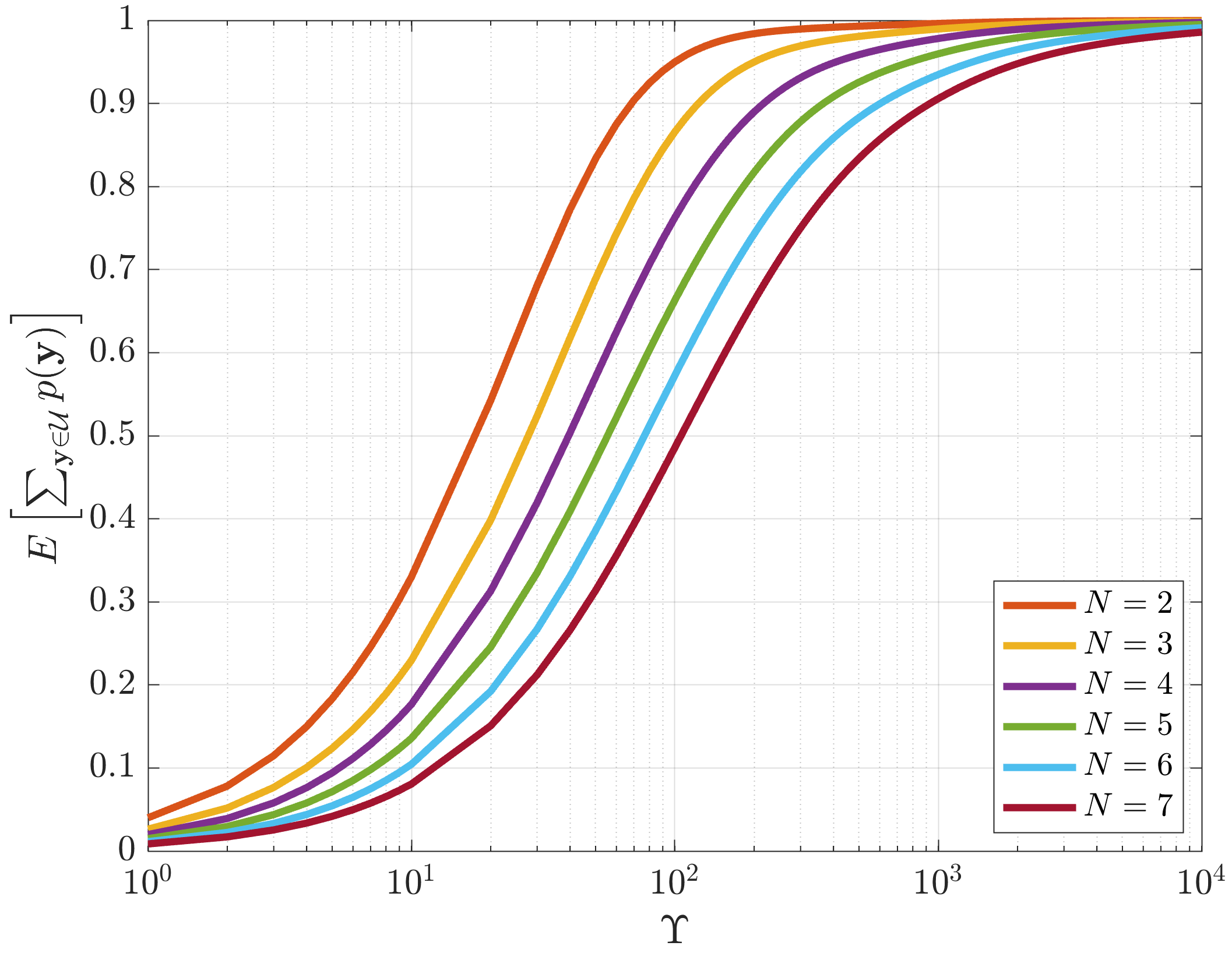}
    \caption{Expected Total Probability Mass (Analogous to $\epsilon$) of All Unique Indexing Sequences Generated by $\Upsilon$ Simulated Samples of Quantized Tone}
    \label{fig:CalYUpsilonSize}
\end{figure}

%A metric analogous to $\epsilon$ for the proposed numerical simulation method is 
The expected total probability mass of all unique indexing sequences encountered in $\Upsilon$ realizations of the input %. This expression
%The expected total probability captured by these sequences 
(see Appendix~\ref{sec:simHPIapprox} for derivation):
\begin{equation}\label{eq:UpsilonExpectation}
    \mathbb{E}\left[\sum_{\mathbf{d} \in \mathcal{U}} p(\mathbf{d})\right] = 1 - \sum_{j=1}^{2^{\beta}} p(\mathbf{d}_{[j]}) \cdot (1-p(\mathbf{d}_{[j]}))^{\Upsilon}
\end{equation}
% Plotting this value as a 
is illustrated in Fig.~\ref{fig:CalYUpsilonSize}  for different $N$. It reveals that even modest values of $\Upsilon$ can very efficiently generate large $\epsilon$ HPI sets.

\subsection{Memory Size:  An Example}
Consider training a LUT for a 3-bit tone with $N=7$ (without bit-masking for simplicity). By inspecting Fig.~\ref{fig:CalYUpsilonSize} we can conclude that generating $\Upsilon = 10^{3}$ samples of $\mathbf{x}$ can produce an HPI set approximation with expected $\epsilon \approx 0.9$. Intuitively $1000$ samples of $\mathbf{x}$ are far more efficient to generate than the requisite $2^{bN} = 2^{21} > 2 \cdot 10^{6}$ evaluations of $p(\mathbf{y} | \boldsymbol{q})$ necessary to compute the HPI set $\mathcal{Y}_{0.9}$ directly. Next we can consult Fig.~\ref{fig:MetricEvaluationOverEps} which shows this $\epsilon$ value is sufficient to achieve  SFDR correction equivalent to the case when $\epsilon=1$ (and within 3 dB of the same MSE improvement) at the high-resolution output, while Fig.~\ref{fig:CalYEpsSize} reveals that this HPI set would require less than $0.01\%$ the total memory size of the $\epsilon=1$ set. Finally, the PSD in Fig.~\ref{fig:HPIPSDComparison} indicates an effectively identical spectral performance of the HPI set after requantization with dithering to fixed-point output precision. 

Even without bit-masking, we can estimate the total memory cost to achieve this 10+ dB SFDR improvement. When using the post-table dithering architecture we can confidently use $\rho=8$ as per Fig.~\ref{fig:MSEoverQRho} since such a value was sufficient even for the higher-fidelity estimate using $N=10$. The size of the HPI set given in Fig.~\ref{fig:CalYEpsSize} for $N=7$ and $\epsilon=0.9$ is $L < 200$. As per (\ref{eq:memorysizedefn}), this gives a total memory requirement of $<1600$ bytes.
By this analysis sequence it should be clear that high-probability indexing is an efficient and powerful tool for LUT memory optimization.

\subsection{Joint Optimization of Memory Parameters}
In Sec.~\ref{sec:memoryopt} we described how LUT memory size can be controlled through $\rho$, $\beta$, and $\epsilon$. Intuitively, reducing memory through any of these parameters implies a trade-off with performance of the LUT as quantified through MSE. The MSE is an appropriate metric for evaluating the LUT at the \verb|estimation| stage in Fig.~\ref{fig:LUTsystemmodel} prior to \verb|dithering| and \verb|requantization|, as the high-resolution estimate ultimately limits the fidelity of the LUT output signal and is stored directly with precision $\rho$ in a post-table dithering architecture as per Sec.~\ref{sec:ditheringarch}. Thus we seek to jointly optimize the memory size of the LUT and the MSE of the LUT output by manipulating these three hyper-parameters. 

\begin{figure}[h]
    \centering
    \includegraphics[width=0.4\textwidth]{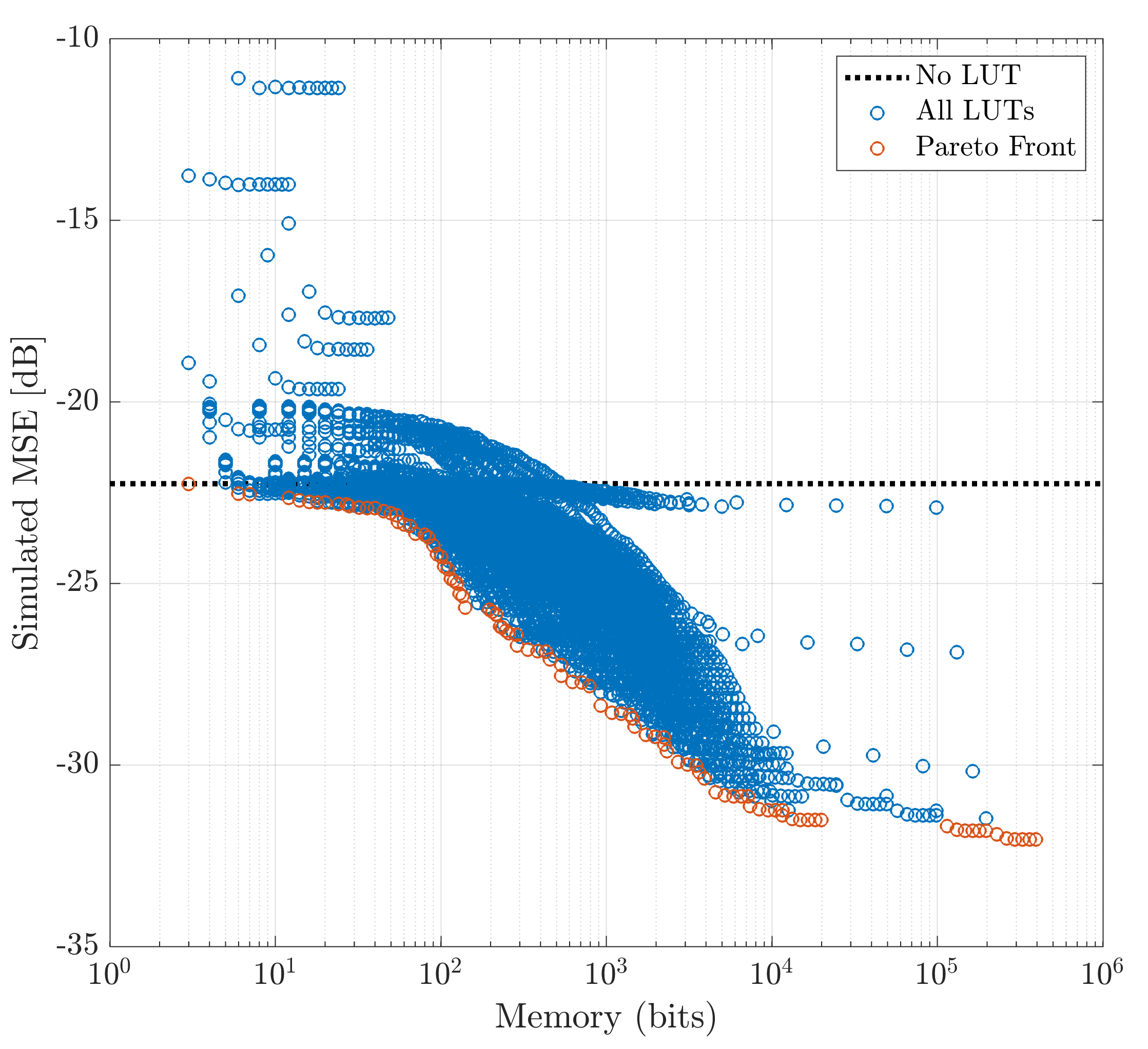}
    \caption{MSE Pareto Front for Simulated Quantized Tone Estimated via Bit-Masked, High-Probability-Indexed LUT and Requantized as $Q_{\rho}(\hat{x}_{\mathrm{MMSE}})$}
    \label{fig:ParetoFront}
\end{figure}

To this end we simulate (with $N=10$) a dense grid of parameters $\beta \in \{1, 2,\cdots, bN/2=15\}$, $\epsilon \in \{0.01, 0.02, \cdots, 1\}$, and $\rho \in \{b=3, 4, \cdots, 12\}$ to determine the Pareto-optimal parameter combinations. For each combination we generate a LUT that stores $Q_{\rho}(\hat{x}_{\mathrm{MMSE}}(\mathbf{d} | \boldsymbol{q}))$ for $\boldsymbol{q}(\beta)$ computed using $H_{3}^{G}$ and $\mathbf{d} \in \mathcal{D}_{\epsilon}(\boldsymbol{q})$. 
%choose $\boldsymbol{q}(\beta)$ according to the Greedy $H_{3}$ heuristic proposed in Sec.~\ref{sec:bitmasksec}, then use this bit-mask to compute $\mathcal{D}_{\epsilon}(\boldsymbol{q}(\beta))$ for  $\epsilon$ from  (\ref{eq:HPIdefn}), then simulate the application of this bit-mask and high-probability indexing to the sinusoidal input described in Sec.~\ref{sec:simsetup} prior to estimation. Lastly, we uniformly quantize the estimator output $Q_{\rho}(\hat{x}_{\mathrm{MMSE}})$ according to resolution $\rho$ for storage in the LUT. 
The result is evaluated for simulated MSE and memory size 
%according to (\ref{eq:memorysizedefn}) 
as $\rho \cdot L = \rho \cdot |\mathcal{D}_{\epsilon}(\boldsymbol{q}(\beta))|$. The resultant dense grid evaluation and Pareto front is shown in Fig.~\ref{fig:ParetoFront}.

\begin{figure}[h]
    \centering
    \includegraphics[width=0.48\textwidth]{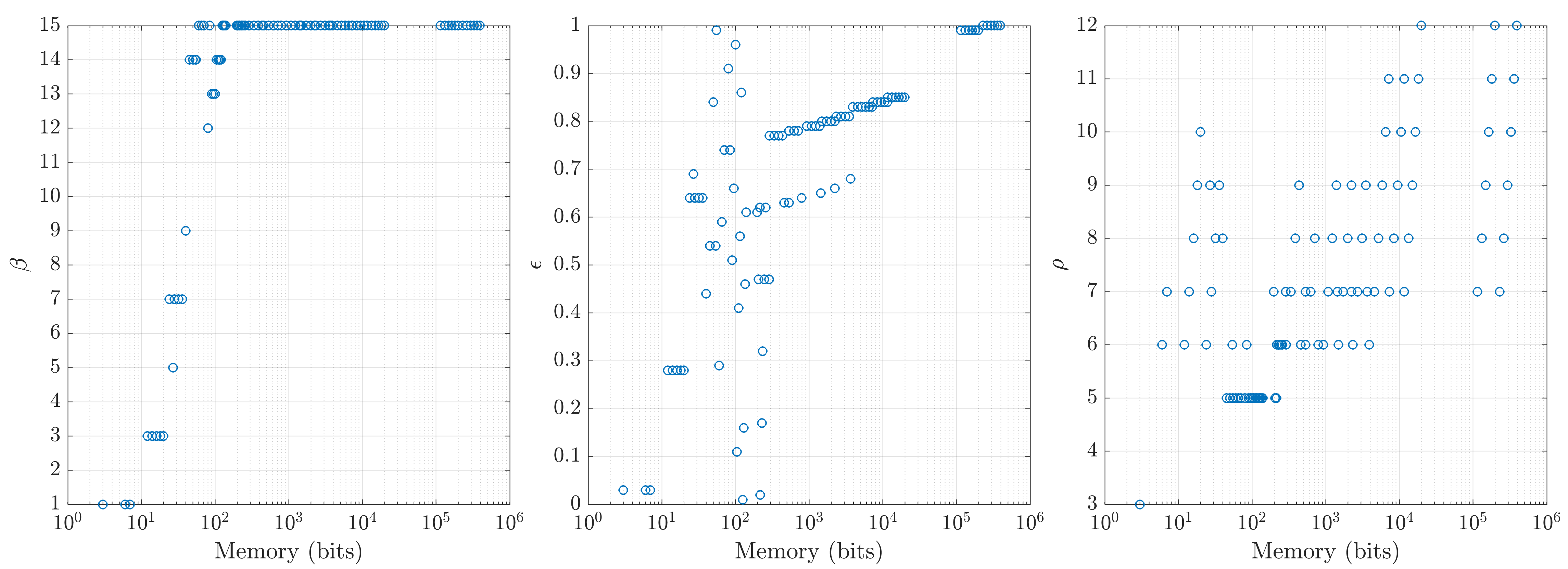}
    \caption{Parameters ($\beta$, $\epsilon$, $\rho$) Producing Pareto-Optimal Points in Fig.~\ref{fig:ParetoFront}}
    \label{fig:ParetoFrontParameters}
\end{figure}

The parameters for points occupying the Pareto front are shown in Fig.~\ref{fig:ParetoFrontParameters}. The main takeaway from this result is that increasing $\beta$ appears to almost always be the most memory-efficient choice, until a maximum $\beta$ is reached per computational limits at which point $\epsilon$ and $\rho$ should be jointly optimized using a grid search. 

\begin{figure}[h]
    \centering
    \includegraphics[width=0.4\textwidth]{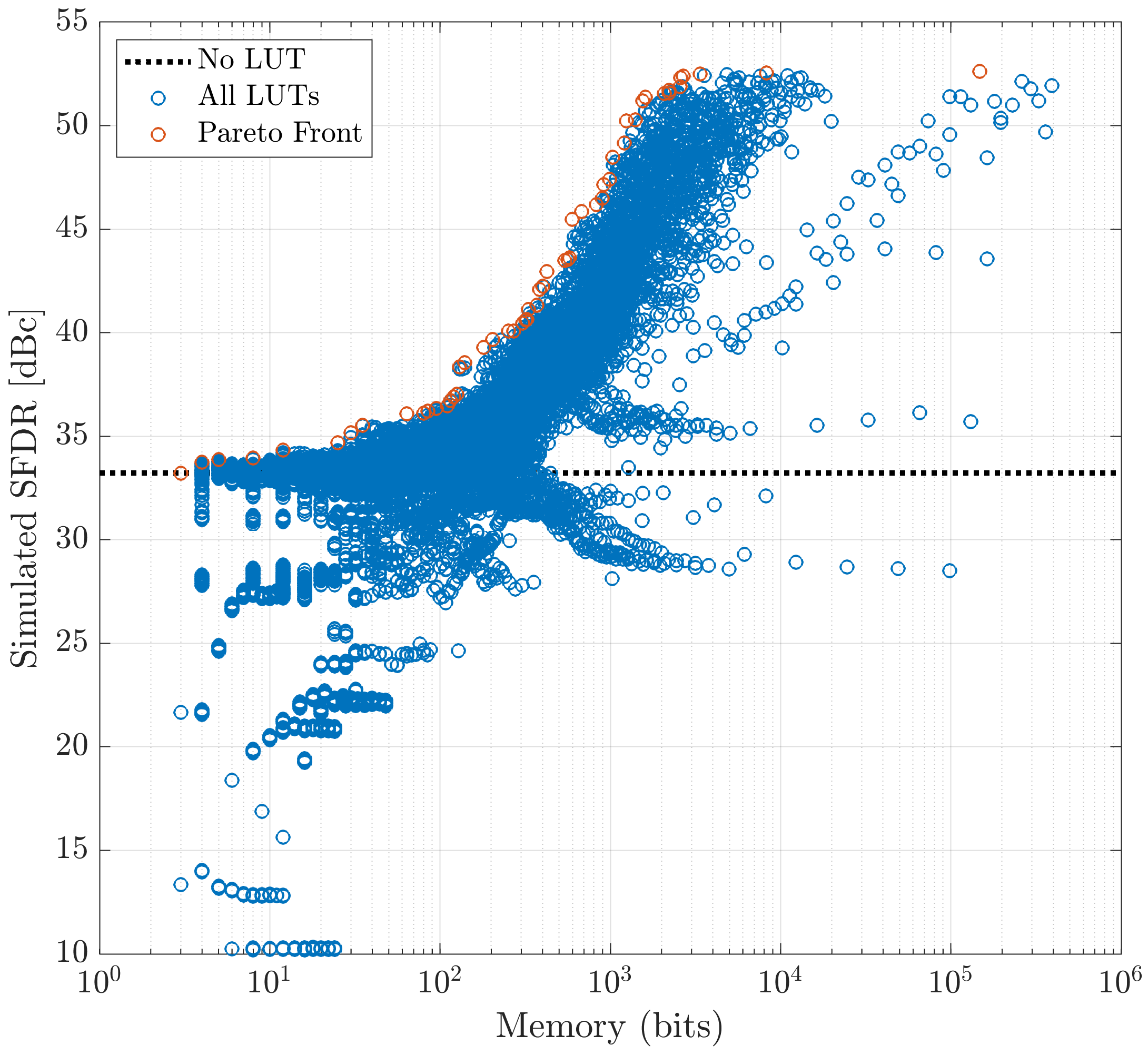}
    \caption{SFDR Pareto Front for Simulated Quantized Tone Estimated via Bit-Masked, High-Probability-Indexed LUT, and Requantized with Post-Table Dithering Using $\alpha=1$ as $Q_{b}(Q_{\rho}(\hat{x}_{\mathrm{MMSE}}) + v)$}
    \label{fig:ParetoFrontSFDR}
\end{figure}

% \begin{figure}[h]
%     \centering
%     \includegraphics[width=0.48\textwidth]{figures/TripleParetoFrontSFDR.png}
%     \caption{Caption}
%     \label{fig:ParetoFront}
% \end{figure}
Next, we study the fixed-point $b=3$-bit output for each of the tested input parameters after post-table dithering using $\alpha=1$. The dense grid evaluation for SFDR and corresponding Pareto front is plotted in Fig.~\ref{fig:ParetoFrontSFDR}. Memory computation is unchanged despite the $b$-bit precision of the output since post-table dithering still requires that the LUT store the estimate with $\rho$-bit precision as described in Sec.~\ref{sec:posttabledithering}.

%Not included is an analysis of SFDR optimization, which necessarily also involves $\alpha$ as per the results in Fig.~\ref{fig:DitheringMethodEvaluation}. However it is expected that the Pareto front for this metric would be even more efficient for a fixed memory constraint. This is because Fig.~\ref{fig:MetricEvaluationOverEps} evidences that even large reductions in $\epsilon$ can maintain the same SFDR despite significant MSE penalties, while $\epsilon$ minimization is among the most potent methods to reduce memory size. 

\section{Conclusion}
We present and evaluate a novel look-up table architecture for real-time all-digital recovery of noisy quantized signals from a parametric input model. The developed system is tested on an example simulated noisy sinusoid quantized to 3 bits. It is proven to be capable of producing a fixed-point digital output that improves the MSE by $>9$ dB while requiring $1446$ bytes of memory to implement. When further constrained to produce an output that maintains the same 3-bit resolution as the input, it is shown to improve the SFDR by $>19$ dBc with only $324$ bytes of memory. The resultant system is thus extremely memory-efficient, low-latency, and compatible with any existing analog system-on-chip by modifying its digital backend using the same total throughput. 

Topics for future work include the study of non-white colored dither generation in post-table dithering, optimal hard-coding of dither sequences for inter-/intra-table dithering, and optimization of the final quantizer stage using Lloyd-Max or alternative methods.
% future work: colored dither for post-table & optimal dither hard-coding for inter-/intra-table dither, quantizer stage optimization (lloyd-max), 

%These performance gains are facilitated by several simultaneous innovations:  One is a detailed study of several proposed post-quantization dithering methods, 

% or


\appendices
\section{$H_{3}$ Derivation}\label{sec:H3derivation}
\begin{align}
    H_{3}(\boldsymbol{q}) &\triangleq \mathbb{E}_{\hat{X}_{0},X_{0}}\left[\left(\hat{X}_{0} - X_{0}\right)^{2} \Big| \boldsymbol{q}\right] \\
    &= \int_{\mathbb{R}} \int_{\mathbb{R}} p(\hat{x}_{0}, x_{0} | \boldsymbol{q}) \cdot (\hat{x}_{0}-x_{0})^{2} d\hat{x}_{0} dx_{0}\nonumber\\
    &= \int_{\mathbb{R}} p(x_{0} | \boldsymbol{q}) \cdot \int_{\mathbb{R}} p(\hat{x}_{0} | x_{0}, \boldsymbol{q}) \cdot (\hat{x}_{0}-x_{0})^{2} d\hat{x}_{0} dx_{0}\nonumber\\
    &= \int_{\mathbb{R}} p(x_{0}) \cdot \sum_{\mathbf{d} \in \mathcal{D}'(x_{0})} p(\mathbf{d}|x_{0},\boldsymbol{q}) \cdot (\hat{x}_{0}(\mathbf{d} | \boldsymbol{q}) -x_{0})^{2} dx_{0}\nonumber\\
    &= \int_{\mathbb{R}} p(x_{0}) \cdot \sum_{\mathbf{d} \in \mathcal{D}'(x_{0})} p(\mathbf{d}|x_{0},\boldsymbol{q})\nonumber\\
    &\hspace{0.5cm} \cdot \left(\frac{\int_{-\infty}^{\infty}x_{0} \cdot p(x_{0}) \cdot p(\mathbf{d} | x_{0}, \boldsymbol{q}) dx_{0}}{\int_{-\infty}^{\infty} p(x_{0}) \cdot p(\mathbf{d} | x_{0}, \boldsymbol{q}) dx_{0}} - x_{0}\right)^{2} dx_{0}\nonumber
\end{align}
where the last step is achieved by substituting (\ref{eq:mmseformulationdv}).

\section{Efficient HPI Generation}\label{sec:simHPIapprox}
When randomly generating $\Upsilon$ indexing sequences, we encounter a set $\mathcal{U}$ of unique indexing sequences $\mathbf{d}_{[j]}$. 
Define the indicator function:
\begin{equation}
    \mathbb{I}_{j} = \left\{\begin{matrix}
        1, &\mathbf{d}_{[j]}\; \mathrm{encountered} \\
        0, &\mathbf{d}_{[j]}\; \mathrm{not\; encountered} \\
    \end{matrix}\right. 
\end{equation}
This allows us to express:
\begin{equation}\label{eq:pIje1}
    p(\mathbf{d}_{[j]} \in \mathcal{U}) = p(\mathbb{I}_{j} = 1)
\end{equation}
Since $\mathcal{U} \subseteq \mathcal{D}$, we have:
\begin{equation}
    \sum_{\mathbf{d} \in \mathcal{U}} p(\mathbf{d}) = \sum_{\mathbf{d}_{[j]} \in \mathcal{D}} p(\mathbf{d}_{[j]}) \cdot \mathbb{I}_{j}
\end{equation}
Taking the expectation and using $|\mathcal{D}| = 2^{\beta}$ with linearity of the expected value operator:
\begin{equation}\label{eq:exppdjexpr}
    \mathbb{E}\left[\sum_{\mathbf{d} \in \mathcal{U}} p(\mathbf{d})\right] = \mathbb{E}\left[\sum_{\mathbf{d}_{[j]} \in \mathcal{D}} p(\mathbf{d}_{[j]}) \cdot \mathbb{I}_{j}\right] = \sum_{j=1}^{2^{\beta}} p(\mathbf{d}_{[j]}) \cdot \mathbb{E}\left[\mathbb{I}_{j}\right]
\end{equation}
Next we apply (\ref{eq:pIje1}) and the assumed independent generation of each of the $\Upsilon$ realizations:
\begin{align}\label{eq:textdj}
    \mathbb{E}\left[\mathbb{I}_{j}\right] &= p(\mathbb{I}_{j} = 1) = p(\mathbf{d}_{[j]} \in \mathcal{U})\nonumber\\
    &= p(\mathbf{d}_{[j]}\; \mathrm{ encountered\; at\; least\; once\; in }\; \Upsilon\;  \mathrm{sequences})\nonumber\\
    &= 1 - p(\mathbf{d}_{[j]}\; \mathrm{ not\; encountered\; in }\; \Upsilon\;  \mathrm{sequences})\nonumber\\
    &= 1 - (p(\mathbf{d}_{[j]}\; \mathrm{not\; realized\; this\; sequence}))^{\Upsilon}\nonumber\\
    &= 1 - (1 - p(\mathbf{d}_{[j]}\; \mathrm{realized\; this\; sequence}))^{\Upsilon}\nonumber\\
    &= 1 - (1 - p(\mathbf{d}_{[j]}))^{\Upsilon}
\end{align}
Substituting (\ref{eq:textdj}) into (\ref{eq:exppdjexpr}):
\begin{align}
    \mathbb{E}\left[\sum_{\mathbf{d} \in \mathcal{U}} p(\mathbf{d})\right] &= \sum_{j=1}^{2^{\beta}} p(\mathbf{d}_{[j]}) \cdot \left(1 - (1 - p(\mathbf{d}_{[j]}))^{\Upsilon}\right)\nonumber\\
    &= \sum_{j=1}^{2^{\beta}} p(\mathbf{d}_{[j]}) - \sum_{j=1}^{2^{\beta}} p(\mathbf{d}_{[j]}) \cdot (1 - p(\mathbf{d}_{[j]}))^{\Upsilon}%\nonumber\\
    %&= 1 - \sum_{j=1}^{2^{\beta}}(1 - p(\mathbf{d}_{j}))^{\Upsilon}
\end{align}
which simplifies into (\ref{eq:UpsilonExpectation}).

% \section*{Acknowledgment}

\ifCLASSOPTIONcaptionsoff
  \newpage
\fi



\bibliography{refs/References}

\begin{thebibliography}{10}
\providecommand{\url}[1]{#1}
\csname url@samestyle\endcsname
\providecommand{\newblock}{\relax}
\providecommand{\bibinfo}[2]{#2}
\providecommand{\BIBentrySTDinterwordspacing}{\spaceskip=0pt\relax}
\providecommand{\BIBentryALTinterwordstretchfactor}{4}
\providecommand{\BIBentryALTinterwordspacing}{\spaceskip=\fontdimen2\font plus
\BIBentryALTinterwordstretchfactor\fontdimen3\font minus \fontdimen4\font\relax}
\providecommand{\BIBforeignlanguage}[2]{{%
\expandafter\ifx\csname l@#1\endcsname\relax
\typeout{** WARNING: IEEEtran.bst: No hyphenation pattern has been}%
\typeout{** loaded for the language `#1'. Using the pattern for}%
\typeout{** the default language instead.}%
\else
\language=\csname l@#1\endcsname
\fi
#2}}
\providecommand{\BIBdecl}{\relax}
\BIBdecl

\bibitem{GrayQuant1998}
R.~Gray and D.~Neuhoff, ``Quantization,'' \emph{IEEE Transactions on Information Theory}, vol.~44, no.~6, pp. 2325--2383, 1998.

\bibitem{GrayDither1993}
R.~Gray and T.~Stockham, ``Dithered quantizers,'' \emph{IEEE Transactions on Information Theory}, vol.~39, no.~3, pp. 805--812, 1993.

\bibitem{wannamaker1997thesis}
R.~A. Wannamaker, ``The theory of dithered quantization,'' 1997.

\bibitem{WannamakerNSD2000}
R.~Wannamaker, S.~Lipshitz, J.~Vanderkooy, and J.~Wright, ``A theory of nonsubtractive dither,'' \emph{IEEE Transactions on Signal Processing}, vol.~48, no.~2, pp. 499--516, 2000.

\bibitem{vanderkooy1984resolution}
J.~Vanderkooy and S.~P. Lipshitz, ``Resolution below the least significant bit in digital systems with dither,'' \emph{journal of the audio engineering society}, vol.~32, no.~3, pp. 106--113, march 1984.

\bibitem{WannamakerPsychoacoustics1992}
R.~A. Wannamaker, ``Dither and noise shaping in audio applications,'' 1992.

\bibitem{lundinthesis}
H.~Lundin, ``Post-correction of analog-to-digital converters,'' Ph.D. dissertation, Royal Institute of Technology (KTH), Stockholm, Sweden, 2003.

\bibitem{lundin_characterization_2005}
H.~F. Lundin, ``\BIBforeignlanguage{en}{Characterization and {Correction} of {Analog}-to-{Digital} {Converters}},'' Doctorate, KTH, 2005.

\bibitem{adctextbook}
H.~Lundin and P.~H{\"a}ndel, \emph{Design, Modeling and Testing of Data Converters}.\hskip 1em plus 0.5em minus 0.4em\relax Springer-Verlag Berlin Heidelberg, 2014, ch. 8 Look-Up Tables, Dithering and Volterra Series for {ADC} Improvements, pp. 249--275.

\bibitem{lundin2003minimalthlut}
H.~Lundin, M.~Skoglund, and P.~H{\"a}ndel, ``Minimal total harmonic distortion post-correction of {ADC}s,'' 2003.

\bibitem{hummels1}
D.~Hummels, F.~Irons, R.~Cook, and I.~Papantonopoulos, ``Characterization of {ADC}s using a non-iterative procedure,'' in \emph{Proceedings of IEEE International Symposium on Circuits and Systems - ISCAS '94}, vol.~2, 1994, pp. 5--8 vol.2.

\bibitem{hummels2}
D.~Hummels, ``Performance improvement of all-digital wide-bandwidth receivers by linearization of {ADC}s and {DAC}s,'' \emph{Measurement}, vol.~31, pp. 35--45, Dec. 2000.

\bibitem{kashersfdrlut}
M.~Kasher, P.~Spasojevic, and M.~Tinston, ``Memory-efficient {SFDR}-optimized post-correction of analog-to-digital converters via frequency-selective look-up tables,'' in \emph{2021 55th Asilomar Conference on Signals, Systems, and Computers}, 2021, pp. 1169--1175.

\bibitem{de_vito_bayesian_2007}
L.~De~Vito, H.~Lundin, and S.~Rapuano, ``Bayesian {Calibration} of a {Lookup} {Table} for {ADC} {Error} {Correction},'' \emph{IEEE Transactions on Instrumentation and Measurement}, vol.~56, no.~3, pp. 873--878, Jun. 2007, conference Name: IEEE Transactions on Instrumentation and Measurement.

\bibitem{lundin_external_2001}
H.~Lundin, M.~Skoglund, and P.~Handel, ``On external calibration of analog-to-digital converters,'' in \emph{Proceedings of the 11th {IEEE} {Signal} {Processing} {Workshop} on {Statistical} {Signal} {Processing} ({Cat}. {No}.{01TH8563})}, Aug. 2001, pp. 377--380.

\bibitem{gines_digital_2021}
A.~Gines, G.~Leger, and E.~Peralias, ``Digital {Non}-{Linearity} {Calibration} for {ADCs} {With} {Redundancy} {Using} a {New} {LUT} {Approach},'' \emph{IEEE Transactions on Circuits and Systems I: Regular Papers}, vol.~68, no.~8, pp. 3197--3210, Aug. 2021, conference Name: IEEE Transactions on Circuits and Systems I: Regular Papers.

\bibitem{attivissimomidpointlinearizationNSdither}
F.~Attivissimo, N.~Giaquinto, A.~Lanzolla, and M.~Savino, ``Effects of midpoint linearization and nonsubtractive dithering in {A/D} converters,'' \emph{Measurement}, vol.~40, pp. 537--544, 06 2007.

\bibitem{lundin_bounds_2009}
\BIBentryALTinterwordspacing
H.~F. Lundin, P.~Händel, and M.~Skoglund, ``Bounds on the performance of analog-to-digital converter look-up table post-correction,'' \emph{Measurement}, vol.~42, no.~8, pp. 1164--1175, Oct. 2009. [Online]. Available: \url{https://www.sciencedirect.com/science/article/pii/S0263224108000365}
\BIBentrySTDinterwordspacing

\bibitem{lundin_adc_2005}
H.~Lundin, M.~Skoglund, and P.~Handel, ``\BIBforeignlanguage{en}{{ADC} {Post}-{Correction} {Using} {Limited} {Resolution} {Correction} {Values}},'' 2005.

\bibitem{lundin_analog--digital_2002}
\BIBentryALTinterwordspacing
H.~Lundin, T.~Andersson, M.~Skoglund, and P.~Händel, ``\BIBforeignlanguage{eng}{Analog-to-{Digital} {Converter} {Error} {Correction} using {Frequency} {Selective} {Tables}},'' Mar. 2002, pp. 487--490. [Online]. Available: \url{https://urn.kb.se/resolve?urn=urn:nbn:se:kth:diva-44275}
\BIBentrySTDinterwordspacing

\bibitem{andersson_frequency_2000}
T.~Andersson, M.~Skoglund, and P.~Händel, ``Frequency estimation by 1-bit quantization and table look-up processing,'' in \emph{2000 10th {European} {Signal} {Processing} {Conference}}, Sep. 2000, pp. 1--4.

\bibitem{kasherfreqestlut}
M.~Kasher, P.~Spasojevic, and M.~Tinston, ``Online memory-constrained frequency estimation for low-resolution non-linear {ADCs},'' in \emph{2022 IEEE Wireless Communications and Networking Conference (WCNC)}, 2022, pp. 956--961.

\bibitem{kasher_postquantization_2024}
M.~Kasher, M.~Tinston, and P.~Spasojevic, ``Post-{Quantization} {Dithering} with {Look}-{Up} {Tables},'' in \emph{2024 58th {Annual} {Conference} on {Information} {Sciences} and {Systems} ({CISS})}, Mar. 2024, pp. 1--6, iSSN: 2837-178X.

\bibitem{widrow_statistical_1996}
B.~Widrow, I.~Kollar, and M.-C. Liu, ``Statistical theory of quantization,'' \emph{IEEE Transactions on Instrumentation and Measurement}, vol.~45, no.~2, pp. 353--361, Apr. 1996.

\bibitem{marco_validity_2005}
D.~Marco and D.~Neuhoff, ``The validity of the additive noise model for uniform scalar quantizers,'' \emph{IEEE Transactions on Information Theory}, vol.~51, no.~5, pp. 1739--1755, May 2005.

\bibitem{lipshitzdithersurvey}
S.~Lipshitz, R.~Wannamaker, and J.~Vanderkooy, ``Quantization and dither: A theoretical survey,'' \emph{Journal of the Audio Engineering Society}, vol.~40, pp. 355--374, 05 1992.

\bibitem{kay1993fundamentals}
S.~Kay, \emph{Fundamentals of Statistical Signal Processing: Estimation Theory}, ser. Fundamentals of Statistical Signal Processing.\hskip 1em plus 0.5em minus 0.4em\relax Prentice-Hall, 1993.

\bibitem{klimesh_quantization_1999}
M.~Klimesh, ``\BIBforeignlanguage{en}{Quantization {Considerations} for {Distortion}-{Controlled} {Data} {Compression}},'' \emph{\BIBforeignlanguage{en}{The Telecommunications and Mission Operations Progress Report}}, vol. 42-139, pp. 1--38, Nov. 1999.

\bibitem{klimesh_optimal_2000}
------, ``Optimal subtractive dither for near-lossless compression,'' in \emph{Proceedings {DCC} 2000. {Data} {Compression} {Conference}}, Mar. 2000, pp. 223--232, iSSN: 1068-0314.

\bibitem{kasher2024dcc_direct}
M.~Kasher, M.~Tinston, and P.~Spasojevic, ``Distortion-controlled dithering with reduced recompression rate,'' in \emph{2024 Data Compression Conference (DCC)}, 2024, pp. 564--564.

\bibitem{kasher2024dcc_arxiv}
\BIBentryALTinterwordspacing
------, ``Distortion-controlled dithering with reduced recompression rate,'' 2024. [Online]. Available: \url{https://arxiv.org/abs/2402.16447}
\BIBentrySTDinterwordspacing

\bibitem{kollar_digital_2006}
I.~Kollar, ``Digital {Non}-{Subtractive} {Dither}: {Necessary} and {Sufficient} {Condition} for {Unbiasedness}, with {Implementation} {Issues},'' in \emph{2006 {IEEE} {Instrumentation} and {Measurement} {Technology} {Conference} {Proceedings}}, Apr. 2006, pp. 140--145, iSSN: 1091-5281.

\bibitem{lundin_criterion_2004}
H.~Lundin, M.~Skoglund, and P.~Handel, ``A criterion for optimizing bit-reduced post-correction of {AD} converters,'' \emph{IEEE Transactions on Instrumentation and Measurement}, vol.~53, no.~4, pp. 1159--1166, Aug. 2004, conference Name: IEEE Transactions on Instrumentation and Measurement.

\bibitem{lundin_optimal_2005}
------, ``Optimal index-bit allocation for dynamic post-correction of analog-to-digital converters,'' \emph{IEEE Transactions on Signal Processing}, vol.~53, no.~2, pp. 660--671, Feb. 2005.

\bibitem{crafts_bayesian_2024}
\BIBentryALTinterwordspacing
E.~S. Crafts, X.~Zhang, and B.~Zhao, ``Bayesian {Cramér}-{Rao} {Bound} {Estimation} with {Score}-{Based} {Models},'' Sep. 2024, arXiv:2309.16076. [Online]. Available: \url{http://arxiv.org/abs/2309.16076}
\BIBentrySTDinterwordspacing

\bibitem{trees_detection_2013}
H.~L.~V. Trees and K.~L. Bell, \emph{\BIBforeignlanguage{en}{Detection {Estimation} and {Modulation} {Theory}, {Part} {I}: {Detection}, {Estimation}, and {Filtering} {Theory}}}.\hskip 1em plus 0.5em minus 0.4em\relax Wiley, Jun. 2013.

\end{thebibliography}
\bibliographystyle{IEEEtran}
\end{document}